\documentclass[printer]{aa} 

\usepackage{graphicx}
\usepackage{txfonts}
\usepackage[]{natbib}
\usepackage{multirow}
\usepackage{array, slashbox}
\usepackage{amsmath}
\usepackage[colorlinks=true,linkcolor=blue,citecolor=blue]{hyperref}
\def\deg{$^{\circ}$~}
\def\degb{^{\circ}}

\begin{document}

 \title{Multiple spiral patterns in the transitional disk of HD\,100546\thanks{Based on data retrieved from the Gemini archive}}
 \author{
 	A. Boccaletti\inst{1},  E. Pantin\inst{2}, A.-M. Lagrange\inst{3}, J.-C. Augereau\inst{3}, H. Meheut\inst{2}., S. P. Quanz\inst{4}
 }
 
 \offprints{A. Boccaletti, \email{anthony.boccaletti@obspm.fr} }

 \institute{LESIA, Observatoire de Paris, CNRS, University Pierre et Marie Curie Paris 6 and University Denis Diderot Paris 7, 5 place Jules Janssen, 92195 Meudon, France
             \and
	        Laboratoire AIM, CEA/DSM-CNRS-Universit? Paris Diderot, IRFU/Service 
d'Astrophysique, CEA/Saclay, 91191 Gif-sur-Yvette Cedex, France
             \and
             Institut de Plan\'etologie et d'Astrophysique de Grenoble, Universit\'e Joseph Fourier, CNRS, BP 53, 38041 Grenoble, France
             \and
	    Institute for Astronomy, ETH Zurich, Wolfgang-Pauli-Strasse 27, 8093 Zurich, Switzerland       
	          }
	   
   \date{}


\authorrunning{A. Boccaletti et al.}
\titlerunning{Multiple spiral patterns in the protoplanetary disk of HD\,100546}

 \abstract
{Protoplanetary disks around young stars harbor many structures related  to planetary formation. Of particular interest, spiral patterns were discovered among several of these disks and are expected to be the sign of gravitational instabilities leading to giant planets formation or gravitational perturbations caused by already existing planets. In this context, the star HD\,100546 presents some specific characteristics with a complex gas and dusty disk including spirals as well as a possible planet in formation. }
{The objective of this study is to analyze high contrast and high angular resolution images of this emblematic system to shed light on critical steps of the planet formation.}
{We retrieved archival images obtained at Gemini in the near IR (Ks band) with the instrument NICI and processed the data using advanced high contrast imaging technique taking advantage of the angular differential imaging. }
{These new images reveal the spiral pattern previously identified with HST with an unprecedented resolution, while the large-scale structure of the disk is mostly erased by the data processing. The single pattern at the southeast in HST images is now resolved into a multi-armed spiral pattern. Using two models of a gravitational perturber orbiting in a gaseous disk we attempted to bring constraints on the characteristics of this perturber assuming each spiral being independent and we derived qualitative conclusions. The non-detection of the northeast spiral pattern observed in HST allows to put a lower limit on the intensity ratio between the two sides of the disk, which if interpreted as forward scattering yields a larger anisotropic scattering than derived in the visible. 
Also, we found that the spirals are likely spatially resolved with a thickness of about $5-10$\,AU. Finally, we did not detect the candidate forming planet recently discovered in the Lp band, with a mass upper limit of $16-18$\,M$_J$.
}
{}
\maketitle

%

\section{Introduction}

\begin{figure*}[ht]
\centerline{
\includegraphics[width=6cm]{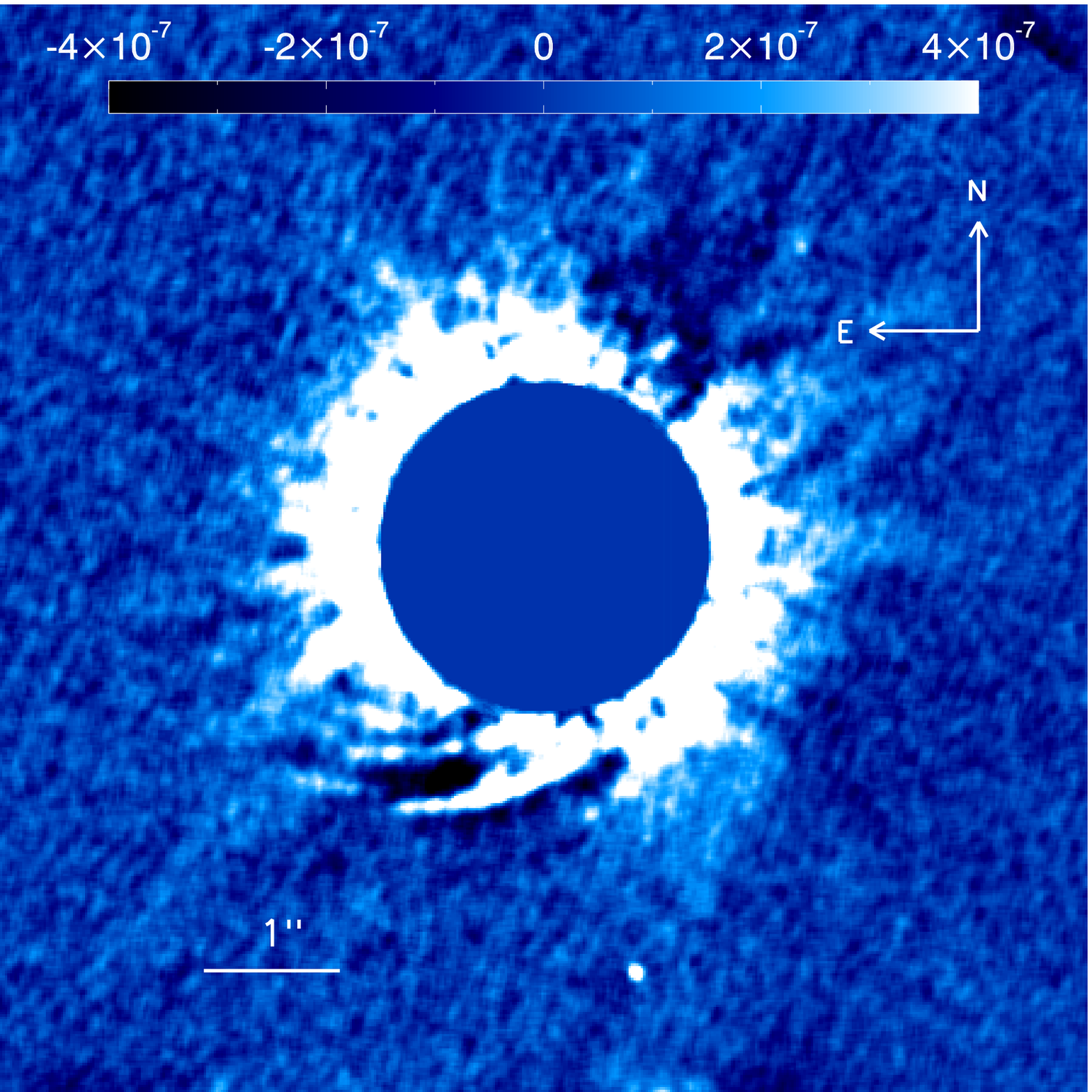}
\includegraphics[width=6cm]{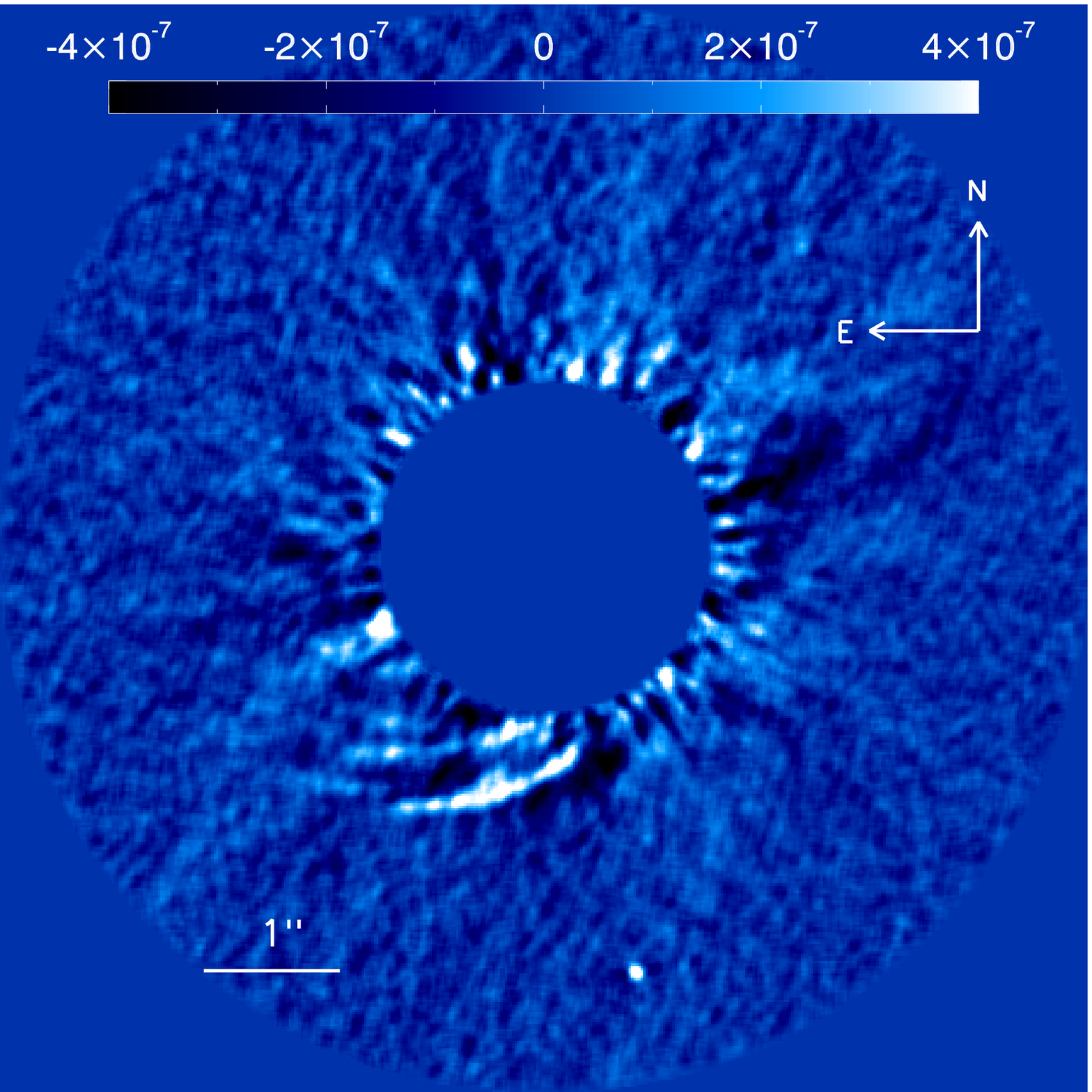}
\includegraphics[width=6cm]{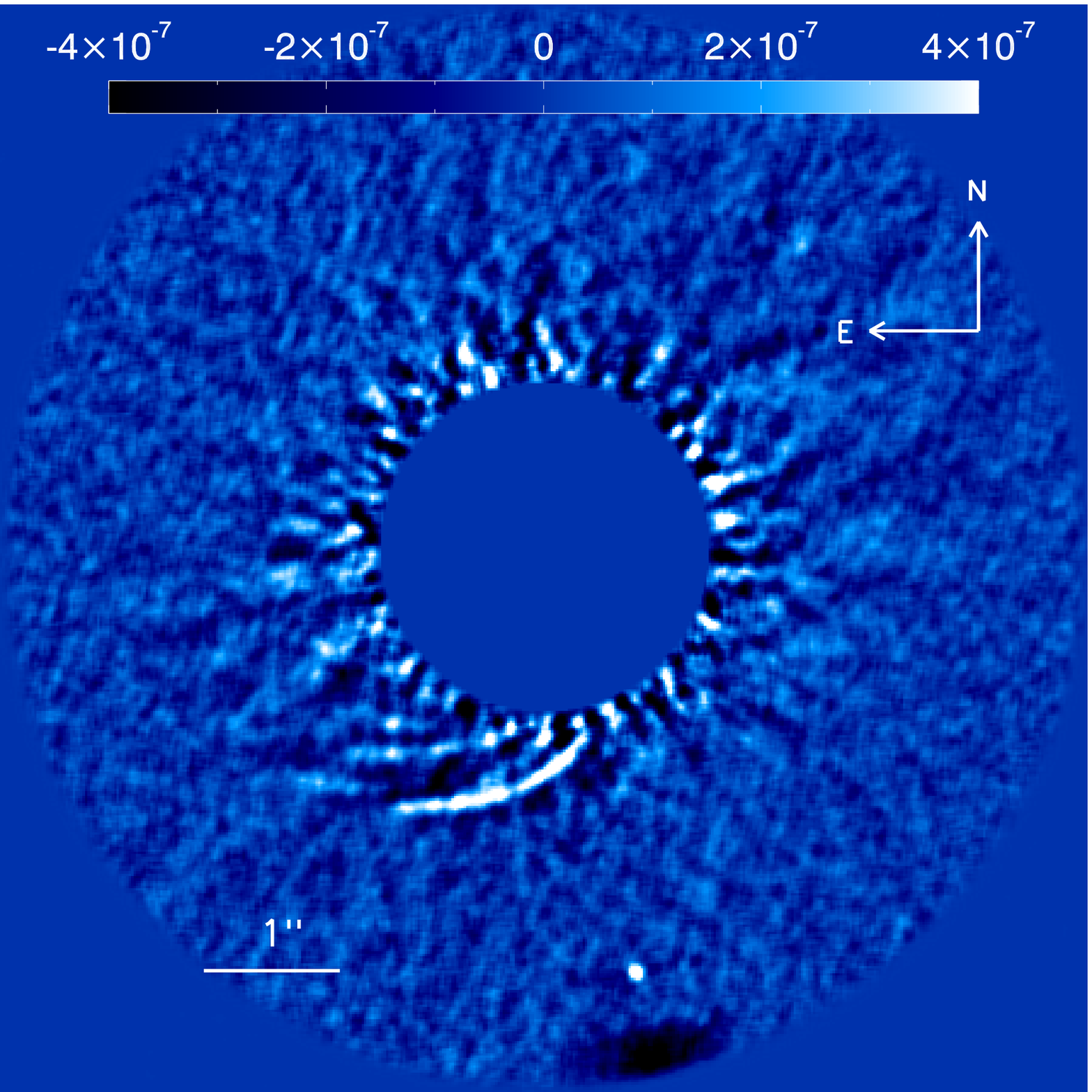}}
\caption{Images of the environment of the star HD100546 as seen in the Ks band with NICI and using several ADI-algorithms: cADI (left), KLIP (middle) and LOCI (right). The field is 8" large and the central area ($<$1.25") is masked out numerically. A 2-pixel smoothing is applied. 
Pixel intensities are coded in contrast with respect to the star.
The point source at the south is corresponding to the object labeled B7 in Fig. \ref{fig:largefield}}

\label{fig:images}
\end{figure*}

Understanding planetary formation requires to identify and describe the many steps of this complex process 
and in particular when they start to form in circumstellar disks.  
 At least two formation mechanisms are proposed to explain the diversity of giant exoplanets as observed today. The so-called core accretion model \citep{Pollack1996} is able to account for the presence of most of the exoplanets discovered by radial velocity and located in the first ten AU from the star. 
At the same time, direct imaging has provided detections of young ($>$10\,Myr) and massive planets  at much larger separations \citep[$>$50\,AU, e.g.][]{Chauvin2005} and, of which the vast majority presumably require gravitational instabilities to form \citep{Boss1998}. 
So far, $\beta$ Pictoris b \citep{Lagrange2010} could be the only imaged exoplanet which may have formed in the regime of core-accretion \citep{Rameau2013}. This peculiar object also indicates that the timescale of formation can be very short ($\leq10$ Myr). However, it is difficult to rely solely on astrometric and photometric measurements in direct images to determine which formation model is at work in a given planetary system. 
Therefore, the observation of younger ($<$10Myr) protoplanetary disks is of prime importance to detect any signs of planetary formation and ideally to catch a planet in formation. In particular, imaging in scattered light allows to map the spatial distribution of the dust on  the disk surface and complement the information collected from unresolved photometric or spectroscopic observations as well as mid infrared  and sub-mm imaging. 

Among the protoplanetary disks in the solar vicinity for which the spatial resolution is fine enough to probe the regions of planetary formation, 
the system HD\,100546 has recently focused a lot of attention with the announcement by \citet{Quanz2013} of a candidate planet in formation. The star is young 
\citep[$5-10$\,Myr,][]{Guimaraes2006} and belongs to the class of Herbig Ae/Be star. However, the large amount of gas observed in absorption and emission \citep{Panic2010, Goto2012} may suggest an age closer to the lower limit of 5\,Myr. 
The presence of a circumstellar disk is suspected since IRAS observations and its spectral energy distribution suggests a warm and a cold grain populations. Silicates emission features comet-like spectrum \citep{Malfait1998}.
Resolved images in the near infrared where first obtained from ground-based Adaptive Optics by  \citet{Pantin2000} and then with NICMOS/HST \citep{Augereau2001}. The disk was seen as an elliptical featureless extended source, which extends far from the star ($\sim$380\,AU) with an inclination of about 51\deg from the line of sight. An asymmetry along the minor axis was marginally detected. To account for the near IR scattered light emission the grains must be larger than 0.1-0.5\,$\muup$m \citep{Mulders2013a}.

Observations in the visible with STIS \citep{Grady2001} and ACS  \citep{Ardila2007} provided an improved angular resolution and sensitivity with respect to near IR and revealed a much more complex morphology. The environment of the star includes a nebulosity extending beyond 8" from the star and a mid-disk (3"-8") where the minor axis asymmetry is interpreted as scattering anisotropy ($g=0.15-0.2$). 
These HST images revealed two trailing spiral arms in the disk region between $1.5-3"$, which develop towards the south-east and the north-west respectively ($150\sim300$\,AU). Moreover, the most detailed images obtained so far shows some hints of additional spiral patterns \citep{Ardila2007}.

The innermost part of the disk, intensively studied by \citet{Benisty2010} and  \citet{Tatulli2011} with AMBER/VLTI, shows an inner hole between 4 and 13\,AU. While the location of the outer rim is confirmed by MIDI/VLTI, \citet{Panic2012} observed that the inner boundary of the gap might be located at 0.7\,AU instead of 4\,AU. The rim at 13\,AU was marginally detected with polarimetric imaging in the near IR  \citep{Quanz2011}. 
Both the spirals and the hole suggest that planetary formation has already occurred in different places of the HD\,100546 environment and interestingly that the two alternative formation mechanisms may coexist in the same system. 
Since then, spiral patterns were observed in other protoplanetary disks \citep{Fukagawa2004, Fukagawa2006, Muto2012, Grady2012, Rameau2012} and are suspected to be the signs of self-gravity and/or planets which triggers spiral wave densities in massive regions of a disk. 

In this paper, we report the observation of HD\,100546 with NICI using public data retrieved from the Gemini archive. The images reveal a set of multiple spiral patterns in the south-east region. Section \ref{sec:obs} presents the observation and the data reduction procedure. The sensitivity to point sources is analyzed in section \ref{sec:point} and the morphology of the spirals is described in section \ref{sec:spiral}. Finally, possible origins of the spirals are discussed in section \ref{sec:discussion}.

\section{Observation and data reduction}
\label{sec:obs}

\begin{table}[ht]
\caption{ Log of observations indicating the date of observation (col. 2), the mask radius (col. 3), the filters (col. 4), the UT at start and at end of the sequence (col. 5), the exposure time of individual frame (col. 6), the amplitude of the parallactic angle variation (col. 7), and the total exposure time integrated (col. 8). }
\begin{center}
\begin{tabular}{lc} \hline\hline
Prog.	& GS-2010A-Q-31 \\
Date			& 2010.03.06 \\
Mask [arcsec]	&0.32	\\ 
 Filters 		&Ks \\
 UT start/end		&04:50:30 / 05:41:40 \\
Exp. time  [s] & 1.9 \\
Parallactic angle  amplitude [deg]	& 17.15 \\
Airmass						& $\sim$1.3 \\
Total exp. time	 [s] & 2215.4 \\ \hline
\end{tabular}
\end{center}
\label{tab:log}
\end{table}

We retrieved archival data obtained with the Near Infrared Coronagraphic Imager \citep[][NICI]{Toomey2003}, installed at  the 8-m telescope, Gemini South. The instrument is based on a near-IR (1-5\,$\muup$m) dual-band imager in which two images are formed simultaneously (owing to a beam splitter) on two separate detectors. NICI has been designed for high contrast imaging and in particular to optimize the search for young planetary mass objects as companion to stars \citep{Liu2010}. In turn, these capacities make NICI suitable for direct imaging of protoplanetary and debris disks as it combines an adaptive optics system, a Lyot coronagraph and angular differential imaging capability \citep{Marois2006}. 

The observation were done on March 6th, 2010 (Tab. \ref{tab:log}) using a single channel configuration in the Ks band (2.15\,$\muup$m) with the 0.32"  flat-topped Gaussian Lyot coronagraphic mask and the 95\% undersized stop. The pixel scale is 18 mas per pixel. The data set consists of 53 files of $1.9s\times22$ coadds $=41.8$ seconds each. The star itself is used for the wavefront sensing. HD\,100546  is a Be star (B9Ve, V=6.7 , K=5.42 ) located at a distance of $97\pm4$\,pc according to \citet{vanLeeuwen2007}. 

We followed the data reduction procedure described in \citep{Boccaletti2013}. The images are dark-subtracted and flat-fielded. The stability of this observing sequence in terms of  AO correction is good enough to retain all the available frames corresponding to a total integration time of $2215.4s$. The registration of coronagraphic images is obtained in two steps, first with a rough estimation of the star position with a Gaussian fit of the images that were thresholded typically at a few percent of the maximum flux \citep{Boccaletti2012}. Thus, frames are registered at the 1-pixel precision. Then, we performed a more precise determination of the star location with Moffat fitting of the central attenuated spot behind the semi-transparent Lyot mask, which is known to move linearly with the actual star position \citep{Lloyd2005}.  A precision of $\sim$0.2 pixel is achieved for the frame registration.

Out-of-mask unsaturated images of the star used to determine the PSF shape and intensity are not available in the archive. Hence, for photometric purpose, we measured the coronagraph attenuation factor separately on a binary star observed with the same settings on May 2010 and we derived a value of 180 ($\sim$5.6$\pm$0.14mag) in agreement with \citet{Wahhaj2011} and \citet{Boccaletti2012}. 
The angular resolution measured on a field star is FWHM$\approx$65 mas.

Then, we processed the data with a set of ADI algorithms: cADI \citep{Marois2006}, LOCI \citep{Lafreniere2007}, and also KLIP, which makes use of a principal component analysis \citep{Soummer2012}.  For a brief description of these algorithms and relevant control parameters see \citet{Lagrange2012b} and  \citet{Boccaletti2012}. Several parameters for these algorithms were tested to achieve the best detection performance. 
The KLIP analysis is applied on an annular region between 20 (the radius of the coronagraphic mask) and 225 pixels (4" in radius) and we retain 1, 2, 3, 4 and 5 modes (out of 53) in the estimation of the stellar contribution. In the LOCI approach, we used the geometry defined by \citet{Lafreniere2007} with $N_A=300$, $g=1$, and $N_\delta=0.5, 1.0, 1.5, 2.0$ in the central 4" region. 
To produce the final images the frames are mean-combined for cADI and KLIP and median-combined for LOCI.
The images presented in Fig. \ref{fig:images} are scaled in contrast with respect to the star maximum intensity measured in the central spot behind the mask in a single pixel. To even further reinforce the speckle attenuation, the KLIP and LOCI images are obtained from the averaging of several reductions with the parameters indicated above.

LOCI produces the sharpest view. A complex set of spiral patterns is detected in the southern part of the field in the region where \citet{Ardila2007} reported the detection of a single spiral (see section \ref{sec:spiral} for a description). 
We note that the range of the parallactic angle along the sequence is relatively small ($17.15\degb$), which is not particularly favorable for the detection of extended objects like circumstellar disks \citep{Milli2012}. 

\begin{figure}[t]
\centerline{
\includegraphics[width=8cm]{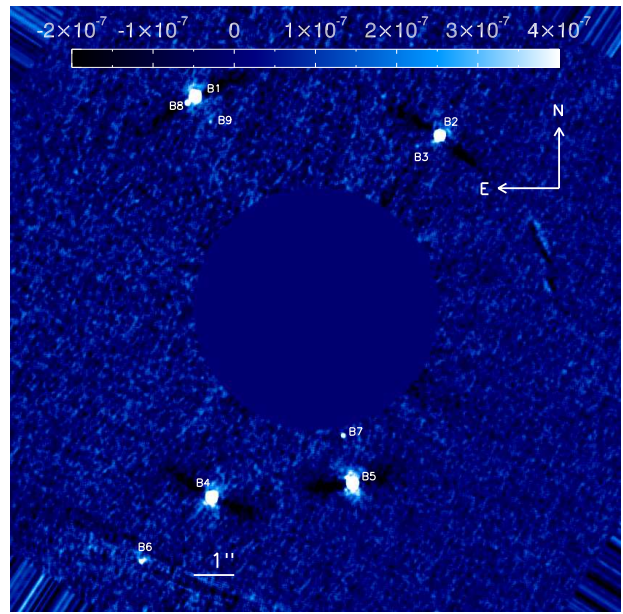}}
\caption{A large 15" field of view image processed with cADI. The central 3" region is masked out. The detected point sources are indicated with labels (B1 to B9). Pixel intensities are coded in contrast with respect to the star.}
\label{fig:largefield}
\end{figure}

\section{Point sources}
\label{sec:point}

\subsection{Background stars}

The 15" field of view displayed in Fig. \ref{fig:largefield} shows many point sources, most of them (B1 to B7) being already identified as background stars by \citet{Ardila2007} based on BVI photometry. A comparison of the NICI image with ACS/HST images in F606W and F814W taken 6 to 7 years apart unambiguously confirms that HD\,100546 has moved with respect to all other sources B1 to B9, as expected from its proper motion ($\mu_\alpha=-38.9$\,mas/yr, $\mu_\delta=0.3$\,mas/yr). 

\subsection{Detection limits}

\begin{table}[t]
\caption{The masses (in unit of Jupiter mass), absolute magnitudes and contrast in the Ks band corresponding to the candidate planet  or several models (COND, DUSTY, BT-SETTL) and two ages (5 and 10 Myr), and assuming all the Lp band flux comes from the planet's photosphere.}
\begin{center}
\begin{tabular}{ccccc} \hline\hline
Age		& Model	& Mass			&	M$_{Ks}$		&	$\Delta$Ks  	\\
(Myr) 	& 		& (M$_J$)			&	(mag)		&	(mag)		\\ \hline
\multirow{3}{*}{5} &	COND	& 	$14-23$	&	$8.2 - 9.4$		&	$7.7-8.9$			\\
& DUSTY	&	$14-23$	&	$8.2 - 9.4$		&	$7.7 - 8.9$			\\
& BT-SETTL&	$14-22$	&	$8.3 - 9.5$		&	$7.8 - 9.0$	\\	\hline
\multirow{3}{*}{10} & COND	& 	$16-27$	&	$8.2 - 9.4$		&	$7.7 - 8.9$		\\
& DUSTY	&	$16-29$	&	$8.2 - 9.4$		&	$7.7 -8.9$		\\
& BT-SETTL&	$16-24$	&	$8.3 - 9.4$		&	$7.8 - 9.0$	\\	
\end{tabular}
\end{center}
\label{tab:model}
\end{table}%

\begin{figure}[t]
\centerline{
\includegraphics[width=9cm]{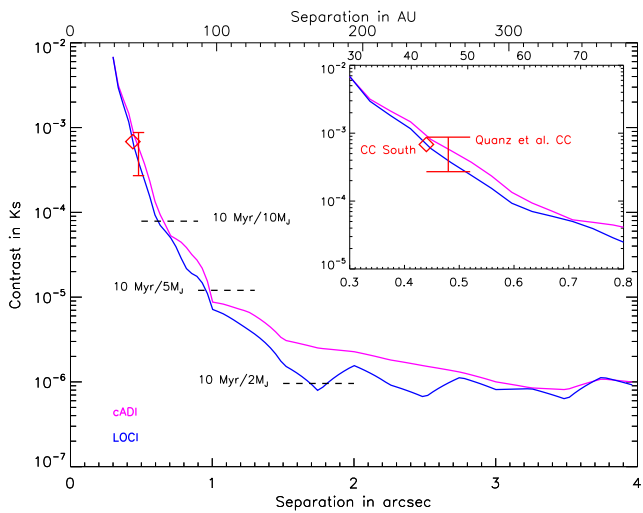}}
\caption{Limit of detection calculated for cADI and LOCI. The red bar shows the expected position in the contrast/separation diagram of the \citet{Quanz2013} candidate planet for which the Ks contrasts are extrapolated from the Lp photometry.
The contrast of a few particular masses (2, 5 and 10\,M$_J$) provided by the BT-SETTL model are plotted for an age of 10 Myr (dashed lines). The red diamond gives the contrast of a candidate companion suspected from the NICI data and located at $\rho=0.44"$, $PA=194.1\degb$. The top-right sub-planel shows a zoom-in version of the same plot.   }
\label{fig:limdet}
\end{figure}

We measured the detection limits in the closer-in region ($<4$") using fake planets injected at known positions/fluxes with respect to the central star. Twenty fake planets are distributed between 0.3" and 3.9" from the star (separated by at least 0.1") in the direction\footnote{Other directions and larger sparsity between the fake planets give identical results.} of the companion candidate reported by \citet[][$\rho=0.48"$, $PA=8.9\degb$]{Quanz2013}. 
We scanned the contrast ratio of these fake objects from $10^{-3}$ to $10^{-7}$ (sampled with 13 values) with respect to the star maximum intensity  and calculated the resulting signal to noise ratio (SNR) of each fake planets.  The signal is taken from the maximum intensity of a fake planet while the noise is the standard deviation of pixels contained on a ring centered at the star with $\pm$ 1 FWHM width and excluding the fake planet.
Then, the SNR is interpolated as a function of contrast (for each fake planets separations) to determine the limit of detection. 
To account for the departure from Gaussian noise in high contrast imaging  \citep{Marois2008}, which here is particularly important as the field rotation is small, we considered a 7 sigma threshold at separations closer than 1.5" while we keep the standard 5 sigma limit further out. These values were determined from a visual inspection of images with fake planets. The resulting detection limits are shown in Fig. \ref{fig:limdet} for cADI and LOCI and were corrected for the coronagraphic mask attenuation as explained in \citet{Boccaletti2013}.
There is no planet more massive than 10M$_J$ (10 Myr, BT-SETTL) at projected separations larger than $\sim$60\,AU (0.62").
 LOCI produces marginally a better starlight suppression than cADI  in the inner 0.7" region while the gain becomes significant further out to approximately 3.5" (about the boundary of the stellar halo).
We overlay the expected contrast in Ks of the putative forming planet discovered by \citet{Quanz2013} with NaCo/VLT assuming a magnitude difference in Lp of $9.0\pm0.5$ mag which translates to an absolute magnitude $M_{Lp}=7.6-8.6$\,mag. We considered several evolutionary models: COND, DUSTY and BT-SETTL  \citep{Chabrier2000, Baraffe2003, Allard2011}, to convert absolute magnitudes to masses from which we derive the expected absolute magnitudes, and then the difference of magnitude in Ks (Tab. \ref{tab:model}). Overall,  the planet/star achieved contrast ratio ranges between $2.6.10^{-4}$ and $8.7.10^{-4}$ ($\Delta$Ks$=7.7 - 9.0$\,mag) rather independently of the age. As a result, the contrast achieved in the NICI data at a separation of 0.48" allows to rule out a self-luminous planet more massive than $16-18$\,M$_J$ according to the age and whatever the evolutionary model. \citet{Quanz2013} suggested that if the planet is still forming the flux measured in Lp may not entirely come from its photosphere and hence we cannot expect the planet intensity to follow the evolutionary models. Our non detection, presented here is still important to set a lower limit, assuming no absorption, on the Ks-Lp color of $1.07\pm0.50$\,mag. More observations will be required to conclude on the companionship of this candidate forming planet (in particular a third epoch) and to compare its spectral energy  distribution with evolutionary models.

\begin{figure}[t]
\centerline{
\includegraphics[width=8cm]{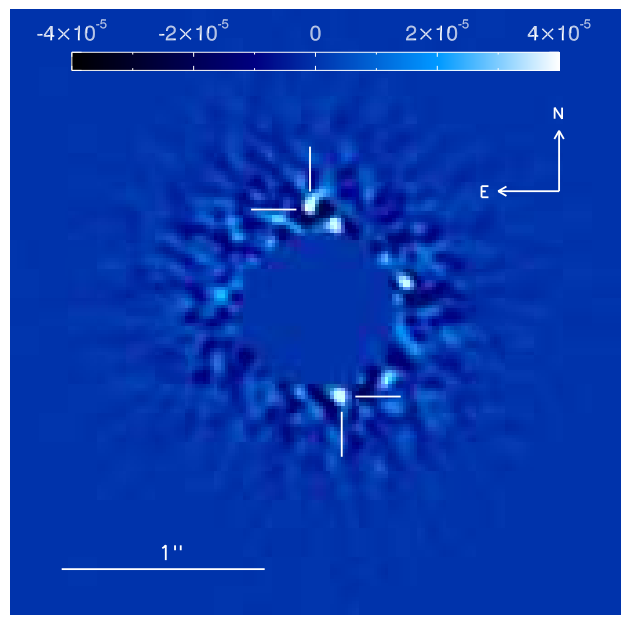}}
\caption{Image of the environment of the star HD100546 as seen in the Ks band with NICI processed with LOCI. The field is 3" large and the central area ($<$0.39") is masked out numerically.}
\label{fig:smallfield}
\end{figure}

\subsection{Additional point sources of interest}

Since planetary formation has presumably started inside the HD\,100546 protoplanetary disk, we searched for other low mass candidates in the field focusing on point-like objects in the star proximity ($<$1"). In Fig. \ref{fig:smallfield} we identified two potential objects, one at the North ($\rho=0.51\pm0.01"$, $PA=4.3\pm0.7\degb$) and one at the South ($\rho=0.44\pm0.01"$, $PA=194.1\pm0.7\degb$). 
The northern one is separated by  $\Delta\alpha=-0.036"$, $\Delta\delta=0.037$ (about two NICI pixels in both axis) from the \citet{Quanz2013} candidate. However, we checked that given the temporal difference of about 15 months, this offset is not consistent with neither the proper motion of the star nor an orbital motion. Moreover, this northern candidate falls right in a region of very bright speckles identified on raw images. Therefore, it is likely an artifact. 

The southern candidate is seen in the LOCI image but also in cADI and KLIP as a more or less extended pattern and at a lower significance. We measured a contrast (corrected from LOCI attenuation) of about $6.8.10^{-4}$ ($\Delta$Ks$\approx7.9$\,mag) after correcting by the mask transmission which is about 40\% at this separation (due to the flat-topped Gaussian profile of the mask). It lies at about our limit of detection so, 7 sigma (Fig. \ref{fig:limdet}), and was not identified by \citet{Quanz2013}. The current data do not allow to conclude on the likelihood of this candidate and more observations will be needed. The relationship between this candidate and the spiral pattern will be discussed in the next section.

\begin{figure*}[t]
\centerline{
\includegraphics[width=8cm]{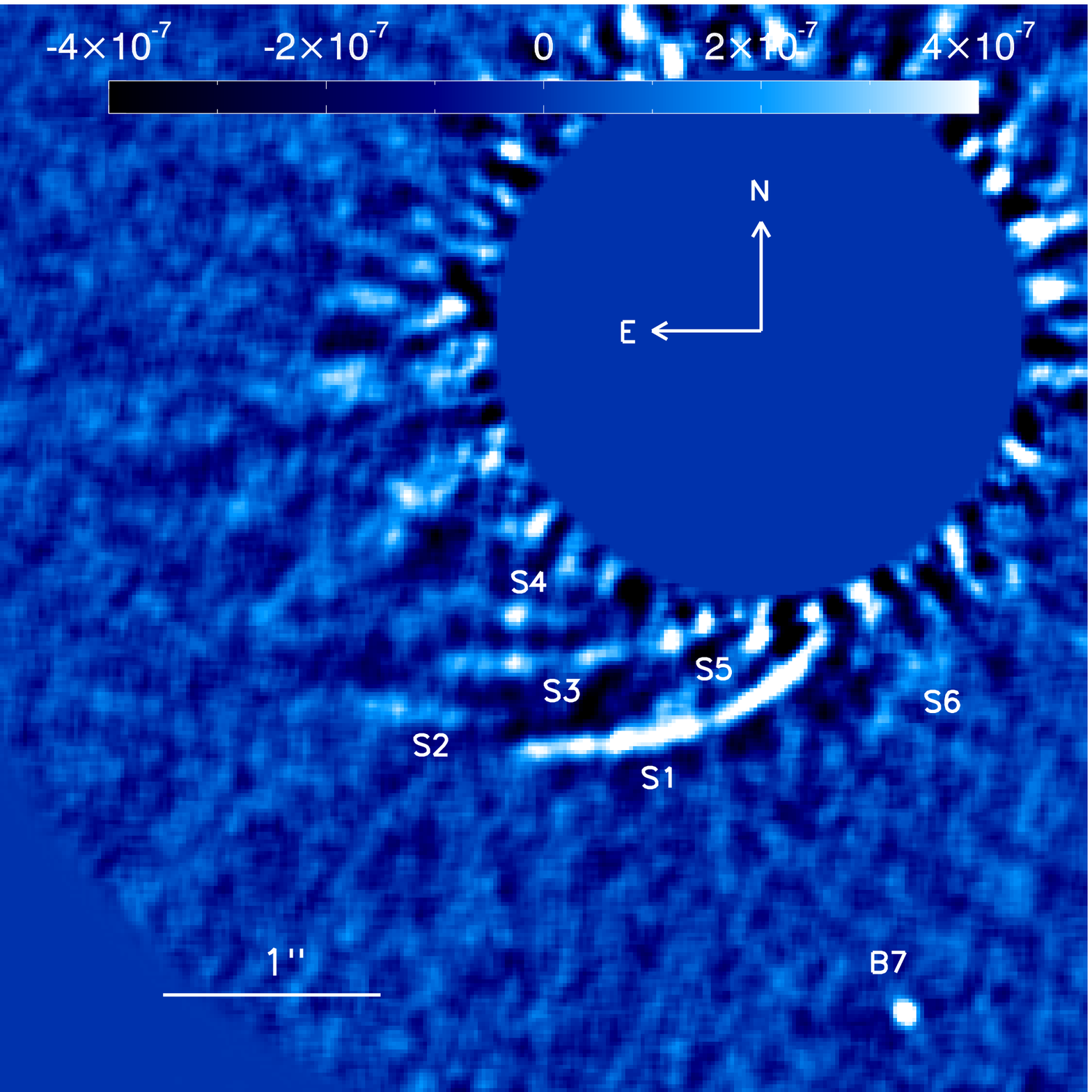}
\includegraphics[width=8cm]{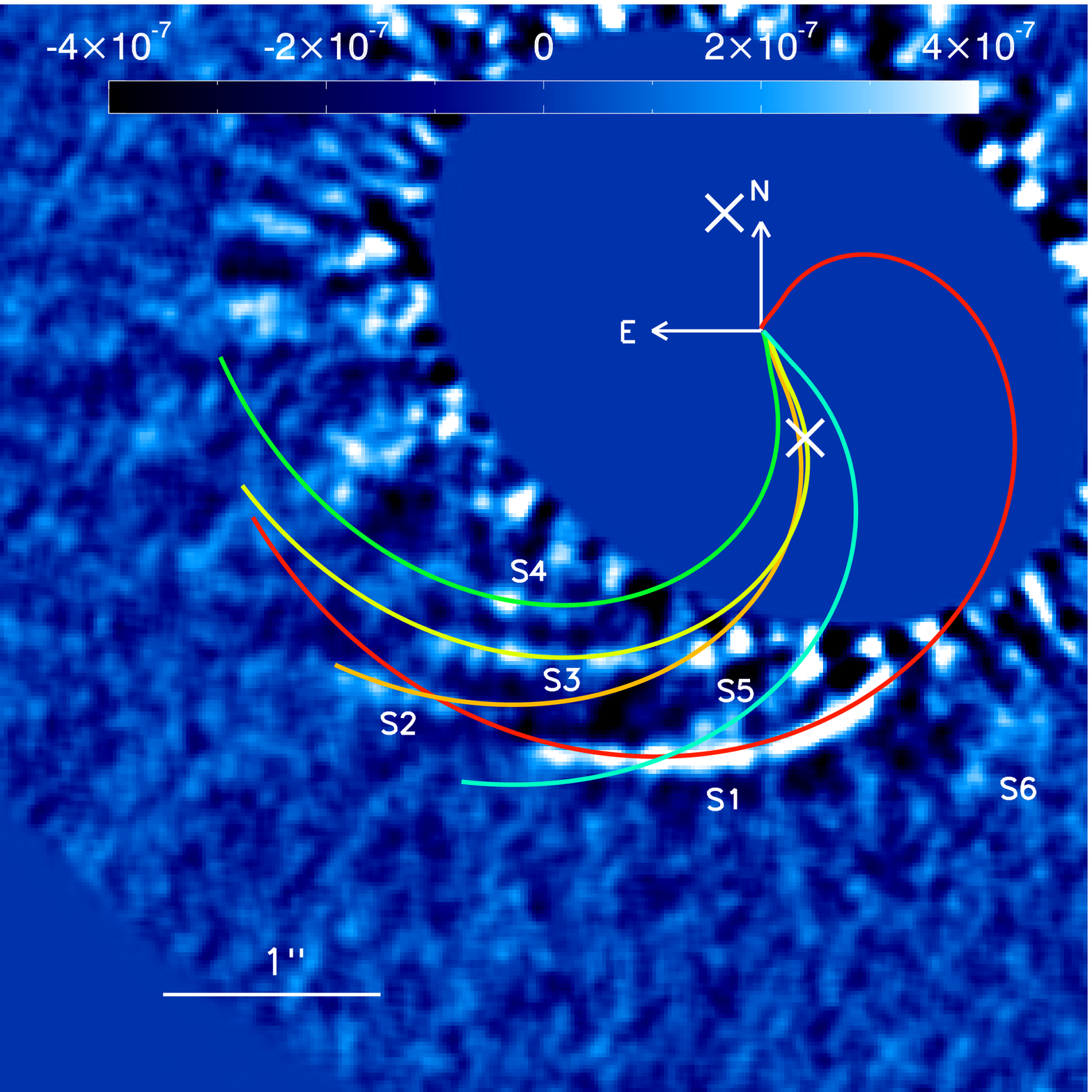}}
\caption{The southern spiral patterns in the LOCI image, with each pieces being labeled (left), and the corresponding fits with a \citet{Muto2012} model calculated on the deprojected image (right). The positions of the \citet{Quanz2013} candidate forming planet and the southern point-source candidate identified in sec. \ref{sec:point} are indicated by two white crosses on the deprojected image.}
\label{fig:zoom}
\end{figure*}

\section{Spiral patterns}
\label{sec:spiral}
\subsection{Description}

\begin{figure*}[t]
\centerline{
\includegraphics[width=8cm]{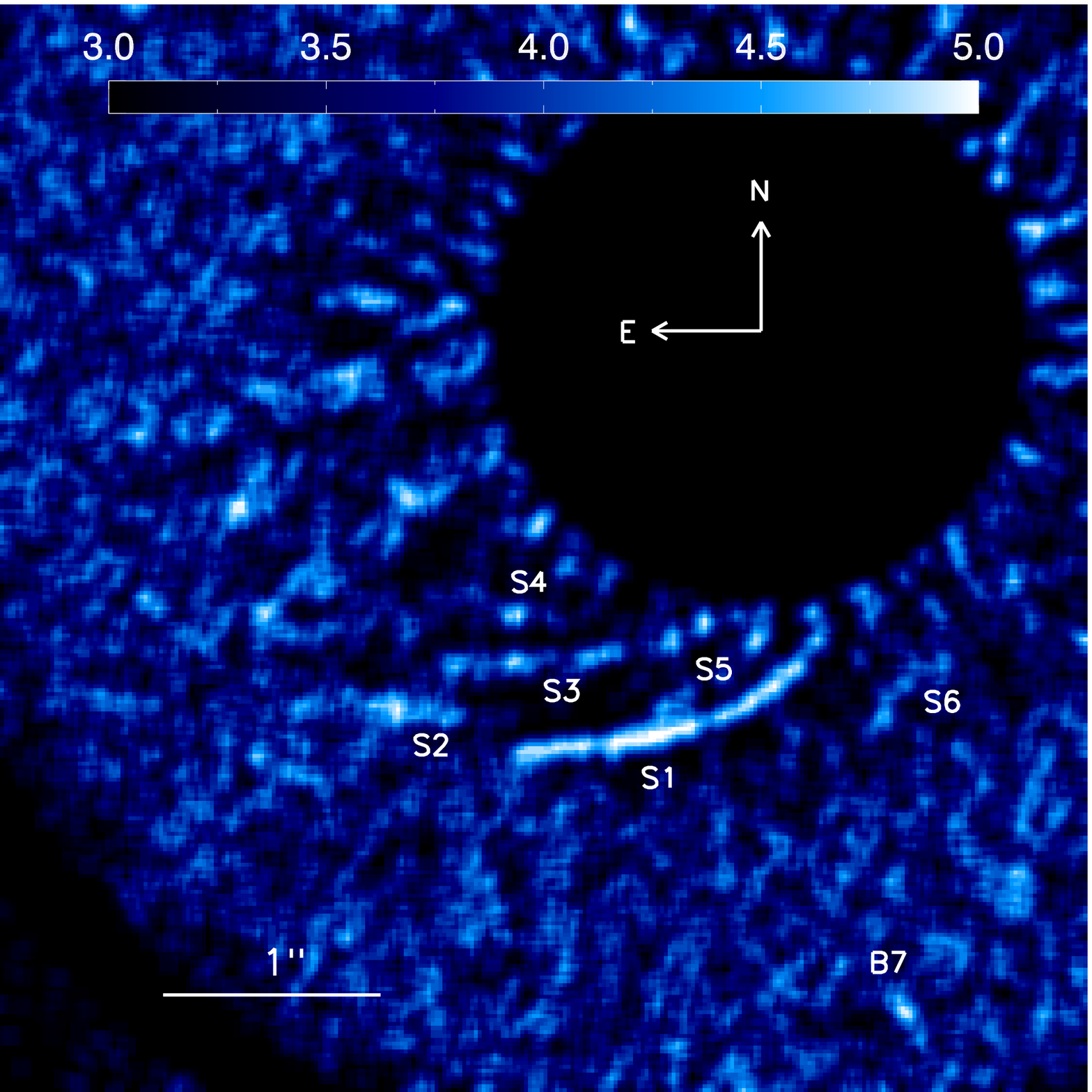}
\includegraphics[width=8cm]{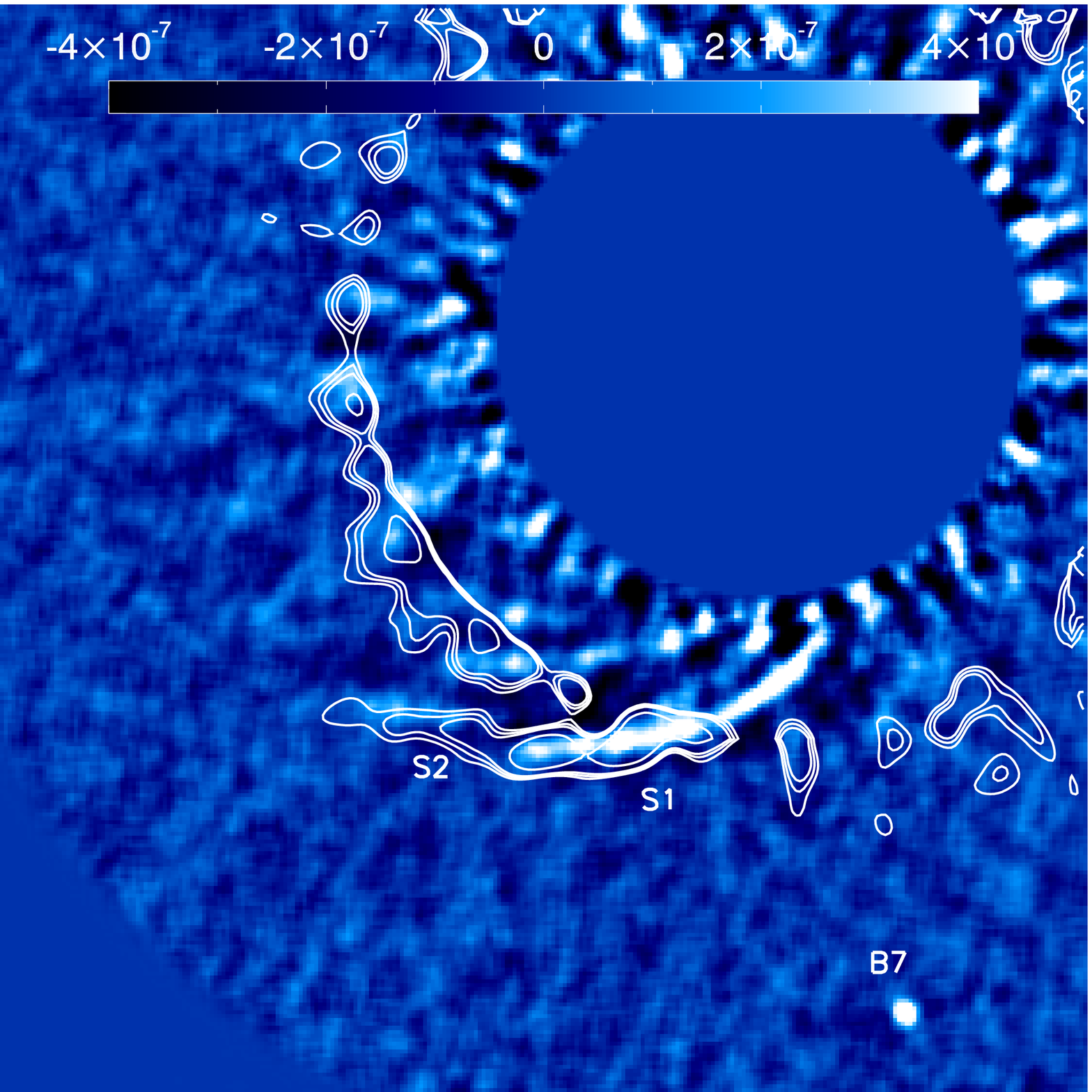}}
\caption{Left: Signal to noise ratio map measured in the LOCI image. S4 and S6 have a better $S/N$ in the cADI image as seen in Fig. \ref{fig:images}. Right: LOCI image as in Fig. \ref{fig:zoom} together with the contour of the F606W ACS image superimposed.}
\label{fig:snrmap}
\end{figure*}

The disk around HD\,100546 is a complex system with a large nebulosity and an inner dusty and gaseous optically thick disk (in the 13-80\,AU region) including some spirals. Here, the observing mode combined with a very small field rotation mostly erase the low-frequency spatial structures in the disk and leave only a set of spiral patterns in the south-east region. We can identify six spiral structures from Fig. \ref{fig:images}, and the cADI process, which is less aggressive than LOCI or KLIP, also shows an extended feature to the southwest ($\rho=2\sim3"$, $PA=200\sim225\degb$ from the North)  with a counterpart in HST images. The most obvious features are labelled S1, S2 and S3 in Fig. \ref{fig:zoom}. The sharpest view of the main spiral patterns is obtained with the LOCI image. 
 S1 is the brightest and longest spiral, it starts from $\rho=1.40"$, $PA= 150\degb$ to $\rho=2.27"$, $PA= 190\degb$. The equivalent projected length is about 155\,AU. No pattern is detected closer in due to the large stellar residuals. Both S1 and S2 are seen in the ACS/HST image as a single broken spiral\footnote{S1 and S2 are noted 3a in Fig. 7 of \citet{Ardila2007}.} but here are identified as two separate features. The inner part of S1, at radius smaller than 1.8", is not seen in the HST image.  S3 is similar to S1 and both are nearly parallel. Three additional patterns S4, S5 and S6 are marginally visible in the LOCI image with S5 lying in between S1 and S3 possibly on a different track.  Both S4 and S6 are well identified in the cADI image in Fig. \ref{fig:images} while they are more attenuated by LOCI. The signal to noise ratio map is shown in Fig. \ref{fig:snrmap} (left) for the LOCI process where the noise is the azimuthal standard deviation for each radius. All spirals have $S/N>3$, locally. This must be considered as a minimum value since the spirals themselves contribute to the azimuthal variations then to the noise.
 A comparison of NICI and HST images obtained 6-7 years apart did not reveal any significant orbital motion of the spirals (Fig. \ref{fig:snrmap}, right). There might be a small offset of S2 but it is unrealistic to conclude on a motion as long as the star position are matching at the 1-pixel level precision, and the field orientation is know to only a few tenths of degree.
All spirals  appear clumpy but it remains delicate to determine what is the impact of the speckle noise and the reduction process in these data, so the surface brightness inhomogeneities along the spirals might be just an artifact. 
We do not detect any spirals in the northern region while S1 has clearly a centro-symmetrical counterpart in the HST data \citep{Ardila2007}  where it appears fainter and thicker. Gas observations indicate the disk rotates counterclockwise and so the spirals would be trailing the rotation \citep{Acke2006}. Assuming photometric asymmetries are due to forward scattering, then the southwest part is the closest to us \citep{Ardila2007}. 

\subsection{Spiral fitting}
\label{sec:spiralfit}

Determining the morphology of the spirals is of prime importance to understand their dynamics and origins.
Among the various mechanisms which can produce spirals in a protoplanetary disk containing a significant amount of gas (see sec. \ref{sec:discussion}), we considered models with gravitational perturbers, where each single perturber produces a single spiral. As an important assumption, we supposed that the dust we observe in scattered light follows the distribution of the gas.
The dynamical models used hereafter are applicable to face-on geometries, so the disk image must be compensated for its inclination. However, the geometrical parameters of the HD\,100546 disk are poorly constrained with near-IR images providing $PA_{nir}=161\pm  5\degb$, $i_{nir}=51\pm3 \degb$  \citep{Augereau2001}, and visible images yielding $PA_{vis}=145\pm5\degb$, $i_{vis}=42\pm5 \degb$  \citep{Ardila2007}. These two sets of values were tested. Qualitatively, the  results are similar but we will present those obtained using $[PA_{vis}$, $i_{vis}]$  as they provide a slightly better matching between models and data.

We started to fit the traces of these spirals with Archimedean relations as suggested by the theoretical work by \citet{Kim2011} where a planet orbit in a gaseous  disk. In such a case, the radius $\rho$ is directly proportional to the azimuth $\theta$ according to the relation:  $\rho=a\theta+b$ in which $a=r_p/ \mathcal{M}_p$, and $b$ is a constant, with $r_p$ and $\mathcal{M}_p$ the orbital distance and Mach number  of the hypothetical planet launching the spiral. Each spiral (S1 to S5) was isolated with a mask in the image and a Gaussian profile is fitted at each $PA$ (relative to the star position) along the spiral to determine its spine. This profile is not locally orthogonal to the spiral so there is a small projection effect. 
From the measured coordinates ($\rho$, $\theta$) of these spines, we fitted the linear relation of \citet{Kim2011} to derive the parameters ($a$, $b$). 
We found that the slopes ($a$) vary from 0.51 to 1.24 depending on spiral, so rather close to unity. S1 and S3, the more extended spirals, are providing the best fits with $a=-0.51\pm0.01$ and $a=-1.04\pm0.07$, respectively. Interestingly, if the Archimedean spirals, but S1, are extended inward they lead to the central star and cross the location of the southern point-like source ($\rho=0.44"$, $PA=194.1\degb$) identified in section \ref{sec:point}. Moreover, if they are extended outward to the northwest they appear at larger physical separations than the northern spirals identified in HST images which would mean that the southern and northern spirals are not connected. This simple analytical study evidences morphological similarities between the different spirals and could suggest a common origin for some of them at least (perturbers). 

Another mathematical framework based on spiral density wave theory is proposed by \citet{Muto2012} to model also the effect of  a gravitational perturber. For a circular planet orbit, the shape of the density wave is given by:
\begin{eqnarray}
 \theta(r) &=& \theta_0-\frac{\text{sgn}(r-r_c)}{h_c}       \nonumber \\
   	&&\times  \left[  \left(\frac{r}{r_c}\right)^{1+\beta} \left\{ \frac{1}{1+\beta}-\frac{1}{1-\alpha+\beta} \left(\frac{r}{r_c}\right)^{-\alpha}  \right\} \right. \nonumber\\
   && \left. -\left( \frac{1}{1+\beta} - \frac{1}{1-\alpha+\beta} \right)\right]
\label{eq:muto}
\end{eqnarray}
with $r_c$, $\theta_0$ the approximate planet location, and $h_c$ the disk aspect ratio at radius $r_c$. It assumes that the disk angular frequency ($\Omega$) and sound speed ($c_s$) vary as power laws with the radius according to $\Omega  \propto r^{-\alpha}$ and $c_s \propto r^{-\beta}$.
To restrain the parameter space we set $\alpha=1.5$ assuming a keplerian rotation, and $\beta=0.25$ following the temperature profile ($T$) measured in the outer disk by \citet{Panic2012} and assuming $T \propto r^{-2\beta}$. 
We still considered the spirals independently and fit Eq. \ref{eq:muto} to each of them (S1 to S5). The parameters of this model are highly degenerated as indicated by \citet{Muto2012} and \citet{Grady2012}. Table \ref{tab:parammuto} summarizes the values of $r_c$, $\theta_0$ and $h_c$ measured for each spiral. The fits of each spiral yield comparable values, which is expectable from the image as all spirals nearly point to the same direction. On average, we obtained $r_c =0.23 \pm 0.04"$, $\theta_0=230\pm 55\degb$, and $hc=0.13 \pm 0.04$, where the error bars corresponds the 1-sigma dispersion of parameters taken from Tab. \ref{tab:parammuto}. 
Changing the initial conditions of the fit or the way the measures are performed (for instance the masking of each single spiral) yields more dispersion. If one or several perturbers are responsible for these spirals, the model of  \citet{Muto2012} derives a physical separation of $20-30$\,AU which is outside of the gap detected from interferometry data \citep{Benisty2010}. 
Moreover, we reached similar conclusions as the former model when the spirals are extended inward and outward.
Interestingly, the fit to S1 appears on a different track than the other fitted spirals. Whether or not it suggests an additional perturber is difficult to conclude, especially given the important degeneracy in between parameters of this model. 
Finally, with the same model, we tried to fit simultaneously several spirals or to force the fit to S1 to cross the position of the potential planet reported by \citet{Quanz2013}, but with no success. 

 We note that a perturber located at $\sim 20$\,AU, will have a period of about 50 years and so the spiral wave, as expected from Eq. \ref{eq:muto}. In such conditions, the spiral pattern in the NICI image should feature a noticeable offset (of the order of 1") with respect to the HST image. Since this is not what we observe, the perturber might be actually further away (of the order of 100\,AU). This inconsistency again illustrates the degeneracies in the model. Indeed, \citet{Muto2012} and \citet{Grady2012} have shown that a family of values ($r_c$, $\theta_0$) can match the spiral shape in a protoplanetary disk.

\begin{table}[t]
\caption{Parameters of the fit to Eq \ref{eq:muto} for each spiral pattern. }
\begin{center}
\begin{tabular}{cccc} \hline\hline
 ID		&	$r_c$ 		&	$\theta_0$ 		&	$h_c$	\\ 
 		&	[arcsec]		&	[deg]				&			\\	\hline
S1		&	0.18			&	326.5			&	0.09	\\
S2		&	0.25			&	202.7			&	0.16	\\
S3		&	0.28			&	206.5			&	0.10	\\
S4		&	0.24			&	192.9			&	0.13	\\
S5		&	0.22			&	223.5			&	0.18	\\
\end{tabular}
\end{center}
\label{tab:parammuto}
\end{table}%

\begin{figure*}[t]
\centerline{
\includegraphics[width=9cm]{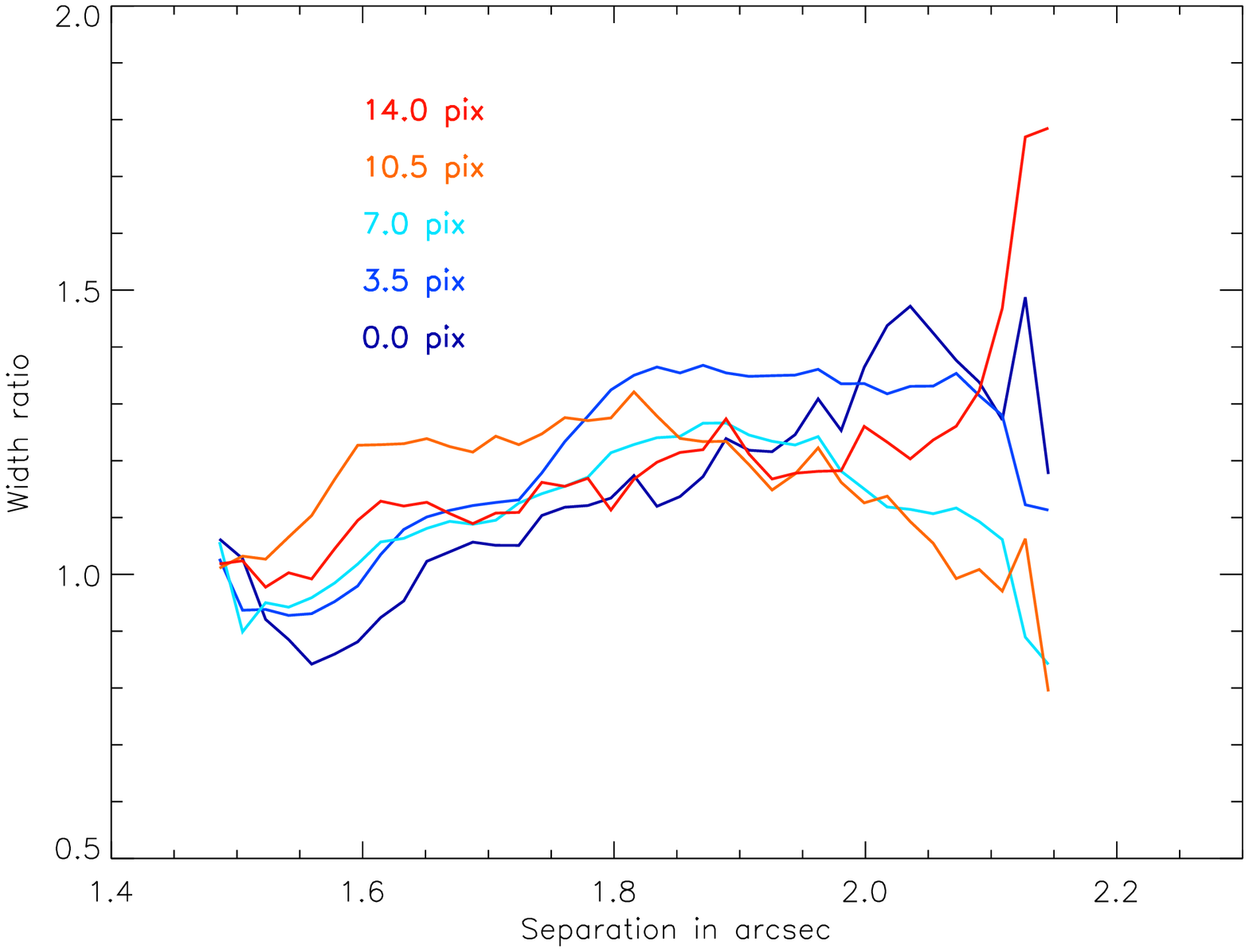}
\includegraphics[width=9cm]{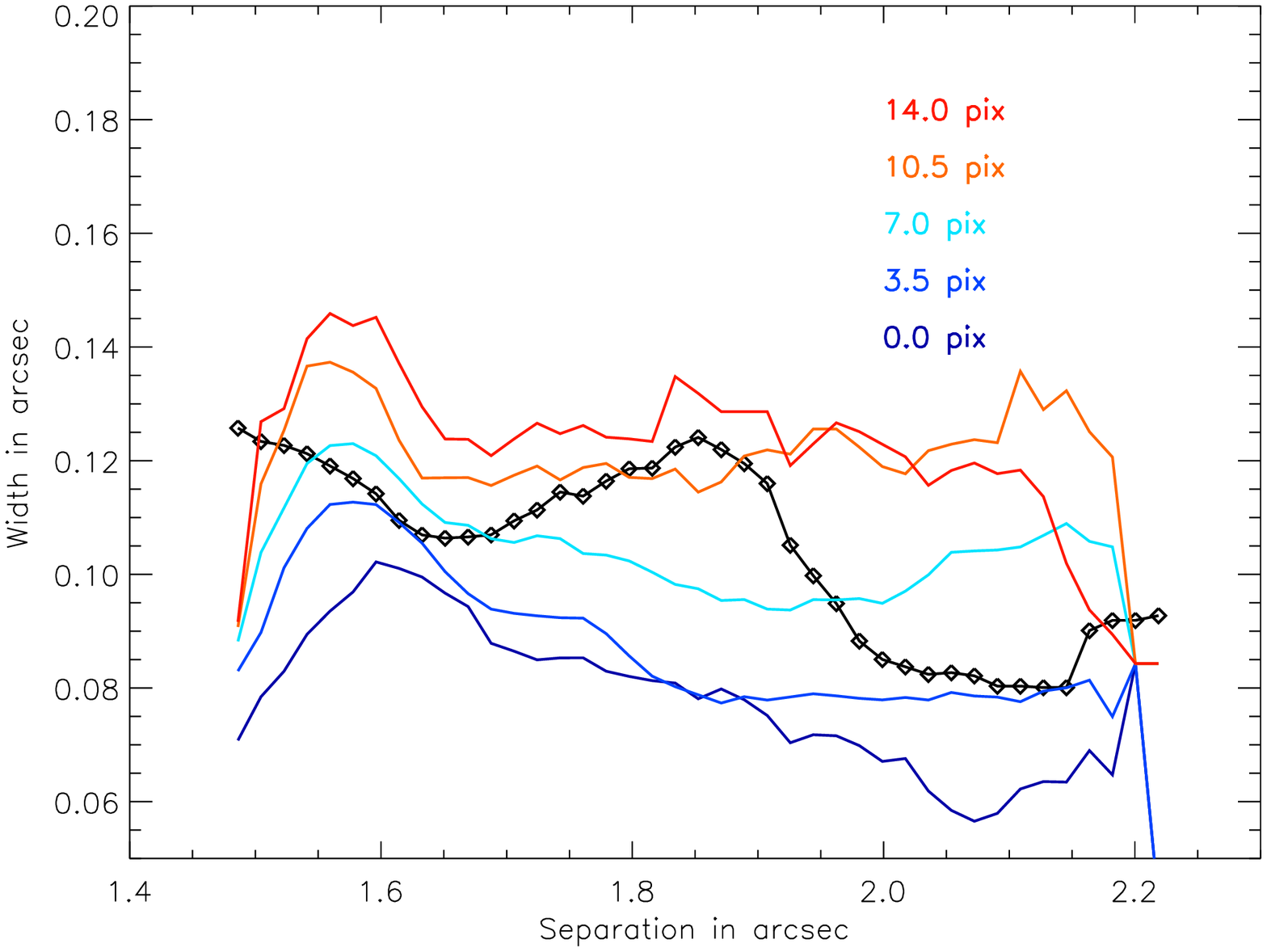}}
\caption{Width ratios (without/with cADI) measured on the fake spirals (left plot), and apparent width (right plot)  measured on S1 (dark line with diamonds), and compared to fake spirals of different sizes (colored lines).}
\label{fig:width}
\end{figure*}

\subsection{Characteristics of the brightest spiral}

\subsubsection{Spatial extension}

To investigate the morphology and photometry of the brightest spiral S1 we generated a fake spiral. We used the coordinates of the fitted \citet{Muto2012} model to the spiral S1 to obtain a trace of the fake spiral.  Convolution with a 2-dimensional Gaussian function produces several fake spirals of different sizes, the FWHM of which is 3.5, 7.0, 10.5, and 14.0 pixels ($\sim$ 65, 125, 190, 250 mas).
The fake spirals, once convolved with the PSF, were implanted symmetrically in the raw data, at 180\deg from the real one, at a contrast of 2.10$^{-6}$ with respect to the star.

We reprocessed the data with the cADI algorithm using an averaged combination of frames in order to better preserve the photometric information. In addition, the fake spiral is also implanted in an empty data cube (no signal, no noise), and the frames are simply derotated and stacked. This provides a reference spiral not affected by any ADI subtraction.
The fake spiral is well detected in the processed image and demonstrates that the small range of parallactic angle quoted in sec. \ref{sec:obs} is not a limitation in fact.

The resultant width was measured in a perpendicular direction with respect to the spine of the spiral. To avoid contamination by other spirals and starlight residuals, S1 was isolated in an annular mask of 10-pixels wide ($\sim$ 0.18"). The spiral profile was fitted with a 1-dimensional Gaussian for each position angle (spanning 40$\degb$) and from a curvature radius located, roughly, at $[-1.06";  0.23"]$ with respect to the star position. The width of the fake spiral located to the northwest was measured the same way.

We first compared the resultant width of the fake spiral measured in the cADI image with respect to that in the empty data cube  
to assess the ADI induced geometrical effect. Figure \ref{fig:width} (left) shows this width ratio as a function of radius (from the star) for several initial sizes of the spiral (3.5 to 14 pix, with 0 corresponding to the unresolved case). The width ratio features a small slope in the 1.6-2.0" range corresponding to a downsizing of 10 to 20\% with respect to the initial width. The apparent width of the spiral S1 is plotted in Fig. \ref{fig:width} (right) against the separation.  The net effect of ADI on the spiral width is not proportional to its width.
The width profile does not decrease monotonically but has a large bump peaking at 1.85". Actually, this radius corresponds to the position where S5 meets S1, so that the width measurement is impacted, inevitably. In addition, the spiral has a dip at about 1.65" which reinforces the appearance of a bump further out. It would not be reliable with the current data to gauge the significance of these features as they might be just artifacts. Nevertheless, from the comparison with fake spirals, we can conclude that S1 is angularly resolved in the direction perpendicular to the spine since it is not coincident with the 0-pixel unresolved case (dark blue line in Fig. \ref{fig:width}). Still, we can determine that the apparent width is about 3.5 to 7 pixels which converts to $\sim$ 6-12\,AU at the distance of HD\,100546.

\subsubsection{Surface Brightness}
\begin{figure}[t]
\centerline{
\includegraphics[width=9cm]{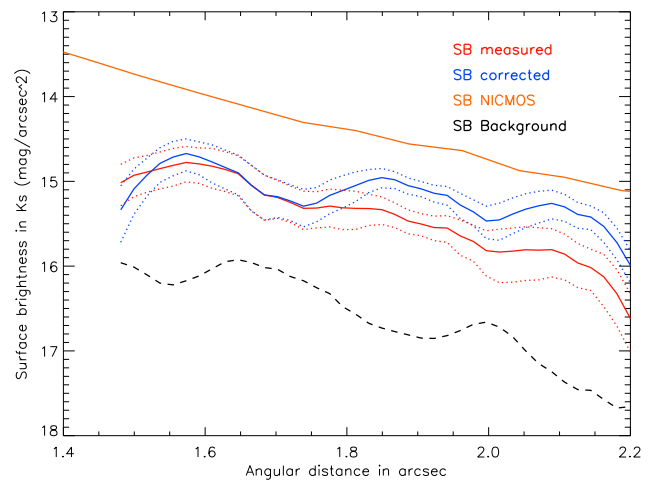}}
\caption{Surface brightness measured in Ks (red line) along S1 and corrected by the ADI attenuation (blue line). The NICMOS surface brightness is taken from \citet{Augereau2001} and scaled in intensity from H to Ks. Dotted lines correspond to the intensity dispersion, while the dashed line} is measured in the background where no spiral is detected.
\label{fig:sb}
\end{figure}

In the following, we considered a 3-pixel wide fake spiral to correct for the ADI photometric artifacts. 
The surface brightness of S1 is measured in the annular mask (10 pixels wide) as defined in the previous section. 
For each position along the spiral, we extracted the averaged intensity and standard deviation perpendicularly to the spiral. The same is performed on the fake spiral with and without the ADI process to calibrate for the flux losses. This correcting factor varies non linearly from about 1 at the inner edge of the spiral (near 1.5") to about 2 at the outer edge (near 2.2"). 
In addition, the averaged intensity of the background is measured at +/- 90\deg from the spiral and averaged. Intensities are converted to magnitude/arcsec$^2$ (using the 2MASS star magnitude and the total flux of the observed PSF), and plotted in Fig. \ref{fig:sb}. As a result, the surface brightness of S1 features a linear decreasing with radius at a rate of $\sim$1 mag/arcsec$^2$/arcsec.  Finally, we compared these measurements to the surface brightness obtained by \citet{Augereau2001} with HST/NICMOS at 1.6\,$\muup$m in the region which encompass the spiral S1. At the first order, and assuming a scattered light regime, the surface brightness from NICMOS is scaled to  2.2\,$\muup$m using the star color (H-Ks=0.54). As expected from ADI, the surface brightness of the spirals is less steep than that of the outer disk (which goes as $r^{-2.9}$) as most of the low frequency components in the image were certainly removed/attenuated by the ADI process. At least, the levels of intensities are consistent with this hypothesis. However, it is difficult to claim the spiral photometry to be accurate here, especially given the small amount of field rotation, but our measurement sets an order of magnitude.

\subsubsection{Anisotropy}

As long as the northern counterpart of the spiral is not detected, this provides a lower limit on the intensity ratio between the southeast and northwest. 
Hence, the scattering properties of the spiral S1 can be inferred from the comparison of its intensity with that of the background. 
As explained in the previous section, we fitted with straight lines the surface brightness (in intensity unit rather than mag/arcsec$^2$) of the spiral S1 (before correction by the ADI attenuation) and that of the background. The intensity ratio varies from 2.7 at the closest separation ($\sim$1.5") to 3.3 at the edge of the spiral ($\sim$2.2"). We assumed $PA_{nir}=161\pm  5\degb$, $i=51_{nir}\pm3 \degb$ as it corresponds to the same spectral range as our data and hence traces the same dust grains. We also assumed that the brightest southwest side of the disk is the closest to us \citep{Ardila2007}. In such conditions, the major axis of the disk  corresponds of 90\deg phase angles while the phase angle of the minor axis are respectively 40\deg in the front (southwest) and 140\deg in the back (northeast). The spiral S1 is located in between the forefront minor and major axes and so spans a range of phase angles. To simplify the calculation, we supposed that S1 is at a median phase angle of  65\deg while that of its symmetrical undetected counterpart would be at 115$\degb$. Assuming Henyey-Greenstein phase function, we derived $g$, the anisotropy scattering factor, which reproduces the measured intensity ratio lower limits. We found a lower limit of $g=0.51_{-0.03}^{+0.10}$, which is larger than the anisotropy scattering factor derived from ACS observations but our measurement is only obtained inside the spirals. At visible wavelengths, in the disk, \citet{Ardila2007} reported $g=0.15 - 0.23$. The two observations might not be directly comparable as we are referring to different phase angles and our measurement is certainly impacted by an un-calibrated ADI attenuation on the low-frequency components of the disk. Still, the numbers are sufficiently different to hypothesize a modification in the scattering properties of the grains from the visible to the near IR. A larger value of $g$ may imply larger grains.
In that respect, \citet{Mulders2013a} has calculated the asymmetry parameter as a function of grain size and showed that $g=0.15-0.23$ in the visible  would correspond to particles of about 0.1\,$\muup$m. But for $g>0.5$, a second set of solutions exist which rather corresponds to a grain size of about 2.5\,$\muup$m (at $\lambda=0.6$\,$\muup$m). Since the particle size scales linearly with wavelength, our measurement of $g$ in the Ks band yields a particle size of about 10\,$\muup$m so a hundred times larger than what inferred from visible images, but measured in a different location. This can be consistent, qualitatively speaking, with the hypothesis that regions of higher pressure, like a spiral (precisely, where we measured $g$), tend to favor grain growth \citep{Rice2006}.

\subsubsection{Scale height}
\label{sec:height}
The fit to spirals in section \ref{sec:spiralfit} yields a disk aspect ratio $h_c=0.09-0.18$.
Assuming spherical particles, composed of silicates \citep{Draine1984}, with a size distribution
following a power-law with -3.5 index, and radii between 1 and 1000 $\mu$m, the dust extinction coefficients can be computed 
using the Mie theory.  Using the disk model parameters obtained by \cite{Benisty2010}, one can estimate the normal optical thickness at a distance of 150\,AU from the star, where the spirals start to emerge from the noise. At wavelengths between 0.4 and 0.8\,$\muup$m, i.e. in the HST/ACS observing
wavelength range, the normal optical thickness is around $\sim$0.6. At NICI observing wavelength (2.12\,$\muup$m), this value remains almost the same.  Therefore, the disk is at the limit between the optically thin/thick cases at these wavelengths and distance from the star. 
Let's suppose the disk to be optically thick, thus, we can apply the ellipse-fitting method developed in \citet{Lagage2006} to measure the disk height.
This method fits elliptical contours using a chi-square minimization based on a Levenberg-Marquardt algorithm.
At a given radial distance of 150\,AU (1.5") from the star, the isophote fit to 
 \citet{Ardila2007} ACS images (and the corresponding offset between the star position
and the ellipse center) implies an estimated scale height of 10$\pm$12\,AU. The error on the scale height is
estimated from the standard deviation of the offsets in the direction along the ellipse main axis
(that in principle should be formally equal to 0.0). Given this relatively large error, only an upper scale height of 22\,AU at 150\,AU from the star
can be deduced from this measurement. This upper limit is compatible with the scale height estimated from spiral-shape fitting ($h=0.13$). On the other hand, the difficulty to measure an effective scale height may be in favor of an optically thin case, which we will consider in the next section.

\section{Discussion}
\label{sec:discussion}

The multiple spirals detected with NICI around HD\,100546 are remarkable in the sense that they are probably the first example of such pattern in a transitional disk (depleted in gas compared to proto-planetary disks). However, it is beyond the scope of this paper to propose a unique explanation to account for these spirals as it will require an important modeling effort and certainly more observations to disentangle between different scenarios. Our objective in this section is simply to present some hypotheses which may eventually guide more careful theoretical works. 

Any large perturbation in a disk of gas tends to launch sound waves resulting in a spiral shape due to the differential rotation of the disk. This shape is given by the rotation and sound speed profiles in the disk. Spirals tend to concentrate the solids that are loosely coupled to the gas. The coupling between gas and solids depend on the size of the solids, and the density and velocity of the gas. With a size of 0.1\,$\muup$m 
measured in the disk of HD\,100546 \citep{Pantin2000,Ardila2007}, these particles should be strongly coupled to the gas and follow the gas density structure, except for the dust situated in the low density region, high above the disk midplane, that will tend to be more concentrated than the gas in the pressure bumps. The dust grains seen in near-infrared are situated in the upper part of the disk, partially coupled to the gas and we can consider that the spiral structures observed are also present in the gas disk. In the following we discuss some plausible scenarios, without being affirmative on which one is the most likely although we can exclude some. 

\begin{itemize}
\item{ {\it Self gravity}: A massive disk can be unstable against gravitational instability that form large amplitude spirals in the gas. The multiple spirals formed by this instability are also distributed around the disk. We assume the Toomre parameter, which measures the degree of instability,  can be approximated by $Q \approx (M_*/M_d)\times h$, where $M_*$ and $M_d$ are mass of the star and the disk respectively, and $h$ is the aspect ratio \citep{Meru2011}. Given the mass of dust \citep[5.10$^{-4} M_\odot$,][]{Benisty2010} and gas \citep[1.10$^{-3} M_\odot$,][]{Thi2011} reported in previous studies for the outer disk ($13-350$\,AU) and the aspect ratio we measured in our work ($h=0.13$), we find $Q =  217 >>1$. We therefore reach the conclusion that the disk is not likely to undergo gravitational instabilities as already noticed earlier by \citet{Quillen2005}. However, we also note that the mass of the gas and the cooling rate are not well constrained while it is fundamental for the determination of the $Q$ parameter. In particular, \citet{Panic2010} found a significant mass of gas at 400\,AU.}

\item{ {\it Warp}: Based on the former conclusion, \citet{Quillen2006} proposed an alternative scenario involving a planet on an inclined orbit (6 to 15$\degb$) causing surface brightness variations in the form of a two-armed spiral provided that the outer disk is optically thin. In this framework, the spirals will be leading instead of trailing features. \citet{Quillen2006} found that a Jovian-mass planet orbiting in the gap cannot be responsible for this warp. On the other hand, the possible planet detected by \citet{Quanz2013}  would satisfy the constraint to produce a tilted warp. However, we note that the simulations presented by \citet{Quillen2006} produce two symmetrical arms which were compliant with the HST images presented in \citet{Ardila2007} but do no longer agree with the NICI data. In particular, the multiplicity of spirals is not reproduced in this model. }

\item{ {\it Rossby instability:} If there is a pressure bump in the disk as the one expected at the border of a planet gap, the disk can become unstable to the Rossby wave instability. This instability has the appealing effect to lift large solid grains to the upper part of the disk, limiting the dust settling as observed in the outer disk of HD\,100546 \citep{Mulders2013a}. However the spiral density waves formed by this instability should also be distributed all around the gas disk.}

\item{ {\it External perturber / fly-by}:  \citet{Quillen2005} have presented hydrodynamic simulations which allow for the formation of spirals caused by the influence of a bound external perturber (planetary to sub-stellar mass range) on a non-self-gravitating gaseous disk.
A qualitative matching with HST images is achieved when tuning the relevant parameters (star/perturber mass ratio, disk aspect ratio, perturber eccentricity). However, they pointed out that no corresponding bound low-mass object is detected in any of the HST data \citep{Augereau2001, Ardila2007}. Our limit of detection clearly confirms this result improving the minimal mass from $\sim$10\,M$_J$ (HST) to $\sim$2\,M$_J$ (NICI), nonetheless in a small field of view. In addition, \citet{Quillen2005}  searched the USNO-B catalog for stellar encounters and found none. As a conclusion, if an external perturber is responsible for the spiral structures it should be a very low mass object like a brown dwarf that encountered the HD\,100546 system a few dynamical timescales earlier.}

\item{ {\it Inner planetary mass perturber}: A planet located inside a disk launches a set of sound waves and due to interference between the waves, a unique spiral forms with a large amplitude. Up to a few spirals of different amplitudes can be formed by a unique companion, but again those would be distributed all around the disk \citep{Kley1999}. In this paper, we modeled the spirals using two independent theories accounting for the presence of a perturber inside a gaseous disk \citep{Kim2011, Muto2012}. Although the agreement is acceptable for each spiral individually, this exercise does not allow to determine the origin of the multiplicity and is not accurate enough to firmly conclude on the perturber properties in a mass/separation parameter space. To explain the presence of the multiple spiral pattern close one to each other with this approach, one would have to assume the presence of multiple objects giving rise to multiple and independent wakes. Signs of planetary mass objects were already reported in the HD\,100546 system: the presence of a gap inner to 13\,AU \citep{Bouwman2003}, a planet in formation at  68\,AU or for explaining the radial structure of the disk \citep{Mulders2013b}.  But in this scenario, there is no connection between the spiral patterns and the grouping of up to five spirals in the same region of the disk is unlikely. }

\item{ {\it Two-face spiral} Assuming the disk to be marginally optically thin as hypothesized in section \ref{sec:height}, there is the interesting possibility that two observed spirals features correspond actually to the same 3D spiral wave entity.  Indeed, in this case, one can detect spiral wave scattered light from both disk sides. 
The projected offset (at a given distance from the star) between these two spiral emissions gives a direct measurement of the disk thickness ($t=o/\sin{i}$, where $t$ is the disk thickness in AU, $o$ the geometrical offset in AU and $i$ the disk inclination assumed to be 42$\degb$).
Applying this method to spirals features S1 and S3 (which are very similar in shape/opening angle), one gets a thickness of $\sim$44\,AU which
corresponds to a disk aspect ratio of $h=0.10-0.15$ in agreement with section \ref{sec:spiralfit}. \cite{Ardila2007} suggest, based on disk angularly resolved colors and asymmetry factor, that the disk frontside has a $PA$ around
235$\degb$. In that case, one would expect in a simple scenario, S3 to be brighter than S1 because of dust absorption. 
We observe the opposite. However, there is no guarantee that the spiral pattern on both disk sides have exactly the same surface
brightness. This surface brightness inversion could proceed from structural inhomogeneities between the 2 sides.
}

\item{ {\it Gas and planetesimals coupling:} If the disk has experienced in the past a self-gravitating phase with formation of spirals, the solids preceding the formation of planetesimals are expected to migrate towards the region of higher pressure and then follow the morphology of the gaseous disk. 
Simulations presented in \citet{Rice2006} show that the planetesimals (and then dust) tend to concentrate in narrow multiple spirals, some being broken patterns. To some extent, these results are qualitatively in accordance with what we observed in the disk of HD\,100546. 
Wether or not the solids can stay in the same spatial distribution long after the  the disk has stopped to be self-gravitating
remains to be investigated.
 }

\end{itemize}

\section{Conclusion}

We have presented high contrast images of the environment of HD\,100546 obtained with NICI/Gemini. The planetary mass object detected by \citep{Quanz2013} in the Lp band, and proposed to be a planet in formation, was not re-detected at shorter wavelengths in the Ks band while the sensitivity we obtained allows to rule out objects more massive than $16-18$\,M$_J$. This confirms that a part of the Lp band flux is not coming from the photosphere of this object. The data reach a contrast limit of $\sim10^{-5}$ and  $10^{-6}$ at about $\sim1"$ and 2" respectively, corresponding to about 5 and 2 masses of Jupiter for the oldest age estimate of 10\,Myr. We also identified a possible point-source to the south, close to the detection limit, which remains to be followed to determine wether or not it is a physical object or an artifact. 

More importantly, our data processing revealed the spiral pattern of HD\,100546 with more contrast and clarity than ever before owing to the observing strategy used in NICI, and resolved this structure, formerly identified with HST in the visible, into a set of multiple spirals at the south-east of the star. The brightest spiral is located at $130-220$\,AU from the star and traceable on a projected distance of 155\,AU. We attempted to fit the traces of these spirals, individually, assuming that they were launched by a planet perturber in a disk of gas. For that, we used two models to derive some constraints on the perturbing bodies, but we note that these parameters are degenerated. More observations and modeling will be definitely needed to derive more quantitative informations on potential perturbers and possible links between the spirals. In addition, the non-detection of the northern spiral observed in HST images is peculiar and led us to derive constraints on grain properties. 

Finally, we draw several hypothesis to account for the presence of these spirals. A self-gravitating disk or the presence of a warp appears unlikely, although the amount of gas in the system is not well constrained. Gravitational perturbations by inner bodies could be the most likely scenario but require several perturbers. Providing the disk is optically thin, an attractive solution would be that a single spiral appears two-folded due to the disk inclination and the reflection of the starlight on each disk surfaces. All these scenarios are to be worked out to identify the most plausible. A modeling effort based on the various available data will be presented in a forthcoming paper. In the end, the origin of the spiral is to be understand with more contrast at closer separation, which next high-contrast facilities will be able to offer.

\begin{acknowledgements}
We would like to thank S. Fromang, S. Charnoz, P. Th\'ebault., F. Meru, R. R. Rafikov and R. Nelson for helpful discussions. In addition, we are grateful to the referee for providing useful comments. 
\end{acknowledgements}

\bibliography{paper_hd100546_biblio.bib}

\begin{thebibliography}{52}
\expandafter\ifx\csname natexlab\endcsname\relax\def\natexlab#1{#1}\fi

\bibitem[{{Acke} \& {van den Ancker}(2006)}]{Acke2006}
{Acke}, B. \& {van den Ancker}, M.~E. 2006, \aap, 449, 267

\bibitem[{{Allard} {et~al.}(2011){Allard}, {Homeier}, \&
  {Freytag}}]{Allard2011}
{Allard}, F., {Homeier}, D., \& {Freytag}, B. 2011, in Astronomical Society of
  the Pacific Conference Series, Vol. 448, Astronomical Society of the Pacific
  Conference Series, ed. C.~{Johns-Krull}, M.~K. {Browning}, \& A.~A. {West},
  91

\bibitem[{Ardila {et~al.}(2007)Ardila, Golimowski, Krist, Clampin, Ford, \&
  Illingworth}]{Ardila2007}
Ardila, D.~R., Golimowski, D.~A., Krist, J.~E., {et~al.} 2007, The
  Astrophysical Journal, 665, 512

\bibitem[{Augereau {et~al.}(2001)Augereau, Lagrange, Mouillet, \&
  M~nard}]{Augereau2001}
Augereau, J.~C., Lagrange, A.~M., Mouillet, D., \& M~nard, F. 2001, Astronomy
  {\&} Astrophysics, 365, 78

\bibitem[{Baraffe {et~al.}(2003)Baraffe, Chabrier, Barman, Allard, \&
  Hauschildt}]{Baraffe2003}
Baraffe, I., Chabrier, G., Barman, T.~S., Allard, F., \& Hauschildt, P.~H.
  2003, Astronomy {\&} Astrophysics, 402, 701

\bibitem[{Benisty {et~al.}(2010)Benisty, Tatulli, Menard, \&
  Swain}]{Benisty2010}
Benisty, M., Tatulli, E., Menard, F., \& Swain, M.~R. 2010, Astronomy {\&}
  Astrophysics, 511, A75

\bibitem[{Boccaletti {et~al.}(2012)Boccaletti, Augereau, Lagrange, Milli,
  Baudoz, Mawet, Mouillet, Lebreton, \& Maire}]{Boccaletti2012}
Boccaletti, A., Augereau, J.~C., Lagrange, A.~M., {et~al.} 2012, Astronomy {\&}
  Astrophysics, 544, 85

\bibitem[{Boccaletti {et~al.}(2013)Boccaletti, Lagrange, Bonnefoy, Galicher, \&
  Chauvin}]{Boccaletti2013}
Boccaletti, A., Lagrange, A.~M., Bonnefoy, M., Galicher, R., \& Chauvin, G.
  2013, Astronomy {\&} Astrophysics, 551, L14

\bibitem[{{Boss}(1998)}]{Boss1998}
{Boss}, A.~P. 1998, \apj, 503, 923

\bibitem[{{Bouwman} {et~al.}(2003){Bouwman}, {de Koter}, {Dominik}, \&
  {Waters}}]{Bouwman2003}
{Bouwman}, J., {de Koter}, A., {Dominik}, C., \& {Waters}, L.~B.~F.~M. 2003,
  \aap, 401, 577

\bibitem[{Chabrier {et~al.}(2000)Chabrier, Baraffe, Allard, \&
  Hauschildt}]{Chabrier2000}
Chabrier, G., Baraffe, I., Allard, F., \& Hauschildt, P. 2000, The
  Astrophysical Journal, 542, 464

\bibitem[{Chauvin {et~al.}(2005)Chauvin, Lagrange, Dumas, Zuckerman, Mouillet,
  Song, Beuzit, \& Lowrance}]{Chauvin2005}
Chauvin, G., Lagrange, A.~M., Dumas, C., {et~al.} 2005, Astronomy {\&}
  Astrophysics, 438, L25

\bibitem[{{Draine} \& {Lee}(1984)}]{Draine1984}
{Draine}, B.~T. \& {Lee}, H.~M. 1984, \apj, 285, 89

\bibitem[{Fukagawa {et~al.}(2004)Fukagawa, Hayashi, Tamura, Itoh, Hayashi,
  Oasa, Takeuchi, Morino, Murakawa, Oya, Yamashita, Suto, Mayama, Naoi, Ishii,
  Pyo, Nishikawa, Takato, Usuda, Ando, Iye, Miyama, \& Kaifu}]{Fukagawa2004}
Fukagawa, M., Hayashi, M., Tamura, M., {et~al.} 2004, The Astrophysical
  Journal, 605, L53

\bibitem[{Fukagawa {et~al.}(2006)Fukagawa, Tamura, Itoh, Kudo, Imaeda, Oasa,
  Hayashi, \& Hayashi}]{Fukagawa2006}
Fukagawa, M., Tamura, M., Itoh, Y., {et~al.} 2006, The Astrophysical Journal,
  636, L153

\bibitem[{{Goto} {et~al.}(2012){Goto}, {van der Plas}, {van den Ancker},
  {Dullemond}, {Carmona}, {Henning}, {Meeus}, {Linz}, \& {Stecklum}}]{Goto2012}
{Goto}, M., {van der Plas}, G., {van den Ancker}, M., {et~al.} 2012, \aap, 539,
  A81

\bibitem[{Grady {et~al.}(2013)Grady, Muto, Hashimoto, Fukagawa, Currie, Biller,
  Thalmann, Sitko, Russell, Wisniewski, Dong, Kwon, Sai, Hornbeck, Schneider,
  Hines, Moro-Mart{\'\i}n, Feldt, Henning, Pott, Bonnefoy, Bouwman, Lacour,
  Mueller, Juhasz, Crida, Chauvin, Andrews, Wilner, Kraus, Dahm, Robitaille,
  Jang-Condell, Abe, Akiyama, Brandner, Brandt, Carson, Egner, Follette, Goto,
  Guyon, Hayano, Hayashi, Hayashi, Hodapp, Ishii, Iye, Janson, Kandori, Knapp,
  Kudo, Kusakabe, Kuzuhara, Mayama, McElwain, Matsuo, Miyama, Morino,
  Nishimura, Pyo, Serabyn, Suto, Suzuki, Takami, Takato, Terada, Tomono,
  Turner, Watanabe, Yamada, Takami, Usuda, \& Tamura}]{Grady2012}
Grady, C.~A., Muto, T., Hashimoto, J., {et~al.} 2013, The Astrophysical
  Journal, 762, 48

\bibitem[{Grady {et~al.}(2001)Grady, Polomski, Henning, Stecklum, Woodgate,
  Telesco, Pi{\~n}a, Gull, Boggess, Bowers, Bruhweiler, Clampin, Danks, Green,
  Heap, Hutchings, Jenkins, Joseph, Kaiser, Kimble, Kraemer, Lindler, Linsky,
  Maran, Moos, Plait, Roesler, Timothy, \& Weistrop}]{Grady2001}
Grady, C.~A., Polomski, E.~F., Henning, T., {et~al.} 2001, The Astronomical
  Journal, 122, 3396

\bibitem[{Guimar{\~a}es {et~al.}(2006)Guimar{\~a}es, Alencar, Corradi, \&
  Vieira}]{Guimaraes2006}
Guimar{\~a}es, M.~M., Alencar, S. H.~P., Corradi, W. J.~B., \& Vieira, S. L.~A.
  2006, Astronomy {\&} Astrophysics, 457, 581

\bibitem[{{Kim}(2011)}]{Kim2011}
{Kim}, H. 2011, \apj, 739, 102

\bibitem[{{Kley}(1999)}]{Kley1999}
{Kley}, W. 1999, \mnras, 303, 696

\bibitem[{{Lafreni{\`e}re} {et~al.}(2007){Lafreni{\`e}re}, {Marois}, {Doyon},
  {Nadeau}, \& {Artigau}}]{Lafreniere2007}
{Lafreni{\`e}re}, D., {Marois}, C., {Doyon}, R., {Nadeau}, D., \& {Artigau},
  {\'E}. 2007, \apj, 660, 770

\bibitem[{{Lagage} {et~al.}(2006){Lagage}, {Doucet}, {Pantin}, {Habart},
  {Duch{\^e}ne}, {M{\'e}nard}, {Pinte}, {Charnoz}, \& {Pel}}]{Lagage2006}
{Lagage}, P.-O., {Doucet}, C., {Pantin}, E., {et~al.} 2006, Science, 314, 621

\bibitem[{{Lagrange} {et~al.}(2012){Lagrange}, {Boccaletti}, {Milli},
  {Chauvin}, {Bonnefoy}, {Mouillet}, {Augereau}, {Girard}, {Lacour}, \&
  {Apai}}]{Lagrange2012b}
{Lagrange}, A.-M., {Boccaletti}, A., {Milli}, J., {et~al.} 2012, \aap, 542, A40

\bibitem[{Lagrange {et~al.}(2010)Lagrange, Bonnefoy, Chauvin, Apai, Ehrenreich,
  Boccaletti, Gratadour, Rouan, Mouillet, Lacour, \& Kasper}]{Lagrange2010}
Lagrange, A.~M., Bonnefoy, M., Chauvin, G., {et~al.} 2010, Science, 329, 57

\bibitem[{Liu {et~al.}(2010)Liu, Wahhaj, Biller, Nielsen, Chun, Close, Ftaclas,
  Hartung, Hayward, Clarke, Reid, Shkolnik, Tecza, Thatte, Alencar, Artymowicz,
  Boss, Burrows, de~Gouveia Dal~Pino, Gregorio-Hetem, Ida, Kuchner, Lin, \&
  Toomey}]{Liu2010}
Liu, M.~C., Wahhaj, Z., Biller, B.~A., {et~al.} 2010, Adaptive Optics Systems
  II. Edited by Ellerbroek, 7736, 53

\bibitem[{Lloyd \& Sivaramakrishnan(2005)}]{Lloyd2005}
Lloyd, J.~P. \& Sivaramakrishnan, A. 2005, The Astrophysical Journal, 621, 1153

\bibitem[{Malfait {et~al.}(1998)Malfait, Waelkens, Waters, Vandenbussche,
  Huygen, \& de~Graauw}]{Malfait1998}
Malfait, K., Waelkens, C., Waters, L. B. F.~M., {et~al.} 1998, Astronomy {\&}
  Astrophysics, 332, L25

\bibitem[{{Marois} {et~al.}(2006){Marois}, {Lafreni{\`e}re}, {Doyon},
  {Macintosh}, \& {Nadeau}}]{Marois2006}
{Marois}, C., {Lafreni{\`e}re}, D., {Doyon}, R., {Macintosh}, B., \& {Nadeau},
  D. 2006, \apj, 641, 556

\bibitem[{Marois {et~al.}(2008)Marois, Lafreni{\`e}re, Macintosh, \&
  Doyon}]{Marois2008}
Marois, C., Lafreni{\`e}re, D., Macintosh, B., \& Doyon, R. 2008, The
  Astrophysical Journal, 673, 647

\bibitem[{{Meru} \& {Bate}(2011)}]{Meru2011}
{Meru}, F. \& {Bate}, M.~R. 2011, \mnras, 410, 559

\bibitem[{Milli {et~al.}(2012)Milli, Mouillet, Lagrange, Boccaletti, Mawet,
  Chauvin, \& Bonnefoy}]{Milli2012}
Milli, J., Mouillet, D., Lagrange, A.~M., {et~al.} 2012, Astronomy {\&}
  Astrophysics, 545, A111

\bibitem[{Mulders {et~al.}(2013{\natexlab{a}})Mulders, Min, Dominik, Debes, \&
  Schneider}]{Mulders2013a}
Mulders, G.~D., Min, M., Dominik, C., Debes, J.~H., \& Schneider, G.
  2013{\natexlab{a}}, Astronomy {\&} Astrophysics, 549, 112

\bibitem[{Mulders {et~al.}(2013{\natexlab{b}})Mulders, Paardekooper, Dominik,
  van Boekel, \& Ratzka}]{Mulders2013b}
Mulders, G.~D., Paardekooper, S.-J., Dominik, C., van Boekel, R., \& Ratzka, T.
  2013{\natexlab{b}}, arXiv.org

\bibitem[{Muto {et~al.}(2012)Muto, Grady, Hashimoto, Fukagawa, Hornbeck, Sitko,
  Russell, Werren, Cur{\'e}, Currie, Ohashi, Okamoto, Momose, Honda, Inutsuka,
  Takeuchi, Dong, Abe, Brandner, Brandt, Carson, Egner, Feldt, Fukue, Goto,
  Guyon, Hayano, Hayashi, Hayashi, Henning, Hodapp, Ishii, Iye, Janson,
  Kandori, Knapp, Kudo, Kusakabe, Kuzuhara, Matsuo, Mayama, McElwain, Miyama,
  Morino, Moro-Mart{\'\i}n, Nishimura, Pyo, Serabyn, Suto, Suzuki, Takami,
  Takato, Terada, Thalmann, Tomono, Turner, Watanabe, Wisniewski, Yamada,
  Takami, Usuda, \& Tamura}]{Muto2012}
Muto, T., Grady, C.~A., Hashimoto, J., {et~al.} 2012, The Astrophysical
  Journal, 748, L22

\bibitem[{Pani{\'c} {et~al.}(2012)Pani{\'c}, Ratzka, Mulders, Dominik, van
  Boekel, Henning, Jaffe, \& Min}]{Panic2012}
Pani{\'c}, O., Ratzka, T., Mulders, G.~D., {et~al.} 2012, arXiv.org, 1203, 6265

\bibitem[{{Pani{\'c}} {et~al.}(2010){Pani{\'c}}, {van Dishoeck}, {Hogerheijde},
  {Belloche}, {G{\"u}sten}, {Boland}, \& {Baryshev}}]{Panic2010}
{Pani{\'c}}, O., {van Dishoeck}, E.~F., {Hogerheijde}, M.~R., {et~al.} 2010,
  \aap, 519, A110

\bibitem[{Pantin {et~al.}(2000)Pantin, Waelkens, \& Lagage}]{Pantin2000}
Pantin, E., Waelkens, C., \& Lagage, P.~O. 2000, Astronomy {\&} Astrophysics,
  361, L9

\bibitem[{{Pollack} {et~al.}(1996){Pollack}, {Hubickyj}, {Bodenheimer},
  {Lissauer}, {Podolak}, \& {Greenzweig}}]{Pollack1996}
{Pollack}, J.~B., {Hubickyj}, O., {Bodenheimer}, P., {et~al.} 1996, \icarus,
  124, 62

\bibitem[{Quanz {et~al.}(2013)Quanz, Amara, Meyer, Kenworthy, Kasper, \&
  Girard}]{Quanz2013}
Quanz, S.~P., Amara, A., Meyer, M.~R., {et~al.} 2013, arXiv.org, astro-ph.GA

\bibitem[{Quanz {et~al.}(2011)Quanz, Schmid, Geissler, Meyer, Henning,
  Brandner, \& Wolf}]{Quanz2011}
Quanz, S.~P., Schmid, H.~M., Geissler, K., {et~al.} 2011, The Astrophysical
  Journal, 738, 23

\bibitem[{Quillen(2006)}]{Quillen2006}
Quillen, A.~C. 2006, The Astrophysical Journal, 640, 1078

\bibitem[{Quillen {et~al.}(2005)Quillen, Varni{\`e}re, Minchev, \&
  Frank}]{Quillen2005}
Quillen, A.~C., Varni{\`e}re, P., Minchev, I., \& Frank, A. 2005, The
  Astronomical Journal, 129, 2481

\bibitem[{Rameau {et~al.}(2013)Rameau, Chauvin, Lagrange, Klahr, Bonnefoy,
  Mordasini, Bonavita, Desidera, Dumas, \& Girard}]{Rameau2013}
Rameau, J., Chauvin, G., Lagrange, A.~M., {et~al.} 2013, arXiv.org, astro-ph.EP

\bibitem[{{Rameau} {et~al.}(2012){Rameau}, {Chauvin}, {Lagrange},
  {Th{\'e}bault}, {Milli}, {Girard}, \& {Bonnefoy}}]{Rameau2012}
{Rameau}, J., {Chauvin}, G., {Lagrange}, A.-M., {et~al.} 2012, \aap, 546, A24

\bibitem[{Rice {et~al.}(2006)Rice, Lodato, Pringle, Armitage, \&
  Bonnell}]{Rice2006}
Rice, W. K.~M., Lodato, G., Pringle, J.~E., Armitage, P.~J., \& Bonnell, I.~A.
  2006, Monthly Notices of the Royal Astronomical Society: Letters, 372, L9

\bibitem[{Soummer {et~al.}(2012)Soummer, Pueyo, \& Larkin}]{Soummer2012}
Soummer, R., Pueyo, L., \& Larkin, J. 2012, arXiv.org, astro-ph.IM

\bibitem[{Tatulli {et~al.}(2011)Tatulli, benisty, Menard, Varni{\`e}re,
  Martin-Za{\"\i}di, Thi, Pinte, Massi, Weigelt, Hofmann, \&
  Petrov}]{Tatulli2011}
Tatulli, E., benisty, M., Menard, F., {et~al.} 2011, Astronomy {\&}
  Astrophysics, 531, 1

\bibitem[{{Thi} {et~al.}(2011){Thi}, {M{\'e}nard}, {Meeus}, {Martin-Za{\"i}di},
  {Woitke}, {Tatulli}, {Benisty}, {Kamp}, {Pascucci}, {Pinte}, {Grady},
  {Brittain}, {White}, {Howard}, {Sandell}, \& {Eiroa}}]{Thi2011}
{Thi}, W.-F., {M{\'e}nard}, F., {Meeus}, G., {et~al.} 2011, \aap, 530, L2

\bibitem[{Toomey \& Ftaclas(2003)}]{Toomey2003}
Toomey, D.~W. \& Ftaclas, C. 2003, Instrument Design and Performance for
  Optical/Infrared Ground-based Telescopes. Edited by Iye, 4841, 889

\bibitem[{{van Leeuwen}(2007)}]{vanLeeuwen2007}
{van Leeuwen}, F. 2007, \aap, 474, 653

\bibitem[{Wahhaj {et~al.}(2011)Wahhaj, Liu, Biller, Clarke, Nielsen, Close,
  Hayward, Mamajek, Cushing, Dupuy, Tecza, Thatte, Chun, Ftaclas, Hartung,
  Reid, Shkolnik, Alencar, Artymowicz, Boss, de~Gouveia Dal~Pino,
  Gregorio-Hetem, Ida, Kuchner, Lin, \& Toomey}]{Wahhaj2011}
Wahhaj, Z., Liu, M.~C., Biller, B.~A., {et~al.} 2011, The Astrophysical
  Journal, 729, 139

\end{thebibliography}
\end{document}